\documentclass[journal]{IEEEtran}

\usepackage{ucs}
\usepackage[utf8x]{inputenc}
\usepackage[cmex10]{amsmath}
\usepackage{cite, amsfonts, amssymb, amsthm, bm, bbm, graphicx, relsize, multirow, booktabs, tikz,subfigure,soul}
\usepackage[american]{babel}
\usepackage[T1]{fontenc}
\usepackage{algorithmic, algorithm}
\usepackage[multiple]{footmisc}
\newtheorem{proposition}{Proposition}
\newtheorem{lemma}{Lemma}

\theoremstyle{definition}

\theoremstyle{remark}

\begin{document}


\title{Energy-Efficient Scheduling and Power Allocation
in Downlink OFDMA Networks  with Base\\ Station Coordination}
\author{ Luca~Venturino,~\IEEEmembership{Member,~IEEE,}
Alessio~Zappone,~\IEEEmembership{Member,~IEEE,}
Chiara Risi,
and~Stefano~Buzzi,~\IEEEmembership{Senior Member,~IEEE}
\thanks{The work of Alessio Zappone has received funding from the German Research Foundation (DFG) project CEMRIN, under grant ZA 747/1-2. The work of Stefano Buzzi and Chiara Risi has been funded by the European Union Seventh Framework Programme (FP7/2007-2013) under grant agreement n. 257740 (Network of Excellence "TREND").}
\thanks{This work was partly presented at the 2013 Future Networks and Mobile Summit, Lisbon, July 2013, and at the 2013 IEEE International Symposium on Personal, Indoor and Mobile Radio Communications, London, United Kingdom, September 2013.}
\thanks{L. Venturino, C. Risi and S. Buzzi are with CNIT and DIEI, University of Cassino and Lazio Meridionale, Cassino (FR), Italy (e-mail: \{l.venturino,chiara.risi,buzzi\}@unicas.it). A. Zappone is with the Dresden University of Technology, Communication Theory Laboratory, TU Dresden, Germany (e-mail: Alessio.Zappone@tu-dresden.de).}
}

\maketitle


\begin{abstract}
This paper addresses the problem of energy-efficient resource allocation in the downlink of a cellular OFDMA system. Three definitions of the energy efficiency are considered for system design, accounting for both the radiated and the circuit power. User scheduling and power allocation are optimized across a cluster of coordinated base stations with a constraint on the maximum transmit power (either per subcarrier or per base station). The asymptotic noise-limited regime is discussed as a special case. 
Results show that the maximization of the energy efficiency is approximately equivalent to the  maximization of the spectral efficiency for small values of the maximum transmit power, while there is a wide range of values of the maximum transmit power for which a moderate reduction of the data rate provides a large saving in terms of dissipated energy. Also,
the performance gap among the considered resource allocation strategies reduces as the out-of-cluster interference increases.

\begin{keywords}
Green communications, energy efficiency, resource allocation, scheduling, power control, cellular network, downlink, base station coordination, OFDMA.
\end{keywords}
\end{abstract}

\section{Introduction}\label{SEC:Introduction}
Orthogonal Frequency Division Multiple Access (OFDMA) is the leading multiaccess technology  in current wireless networks, mainly due to its ability to combat the effects of multipath fading \cite{Goldsmith_book}. In order to increase the capacity of OFDMA-based networks, attention has been devoted to the derivation of adaptive resource allocation schemes, which take into account factors such as traffic load, channel condition, and service quality. In particular, base station (BS) coordination has emerged as an effective strategy  to mitigate downlink co-channel interference. Assuming that the data symbols are known only by the serving BS, several papers have shown that joint scheduling and power control among a set of coordinated BSs based on channel quality measurements can greatly improve the network sum-rate \cite{Das-Viswanathan,venturino-2007,Gesbert-2007-12,Gesbert-2008-01,venturino-2008,Gesbert-2008-08,venturino-2009,Honghai-2011,FTN-sum-rate-review-2012}.

On the other hand, environmental and economic concerns require to also account for the energy efficiency of a data network \cite{yu2012green}.
This topic has recently gained big momentum and a number of special issues, conferences, and research projects have been devoted to \emph{green communications} in the last few years: see, for instance, \cite{EARTH,JSACspecialissue,CMspecialissue}, as the tip of the iceberg. Energy-aware design and planning is
motivated by the fact that wireless networks are responsible of a fraction between 0.2 and 0.4 percent of total carbon dioxide emissions \cite{globalfootprint2011}, and this value is expected to grow due to the ever-increasing number
of subscribers. Indeed, energy efficiency will be a key issue also in future fifth-generation cellular networks \cite{whatwillbe}.
The biggest efforts to increase the energy efficiency of a wireless network are concentrated on the access network, since it consumes the largest portion of energy \cite{howmuch2011}.
Here, potential solutions include energy-saving algorithms for switching on and off BSs that are either inactive or very lightly loaded \cite{cellzooming}, energy-efficient hardware solutions \cite{muting}, and energy-efficient  resource allocation algorithms.
Focusing on this latter issue, in \cite{goodman2000power,saraydar2001pricing,buzzi2009noncooperative,buzzi2010transmitter,belmega2011energy}, the energy efficiency is defined as the ratio between the throughput and the transmit power, and the transmit power level maximizing the amount of data bits successfully delivered to the receiver for each energy unit is derived. A more general definition of the energy efficiency is obtained when the circuit power dissipated to operate the devices is included as an additive constant at the denominator. This approach has been considered in \cite{Betz2008}, where power control for direct-sequence code division multiple access multiuser networks is tackled, in \cite{Chong2011c}, where energy efficient communication in a single-user  multiple-input multiple-output (MIMO) system is studied, and in \cite{ZapponeTWC}, where power control in relay-assisted wireless networks is considered. In \cite{circuit_power_consumption,Isheden2011}, several models for the circuit power consumption in wireless networks are elaborated. 
In \cite{miao2010energy},  the tradeoff between energy- and spectral-efficient transmission in multicarrier systems is investigated. The papers \cite{miao2012low,ZapOFDMA} focus on the uplink of an OFDMA system; the former considers a single-cell system and derives low-complexity scheduling and power control strategies, while the latter uses a game-theoretic approach to derive decentralized resource allocation strategies for a multi-cell OFDMA system. With regard to the downlink of an OFDMA system, recent contributions include \cite{limited,large}. The paper \cite{limited} uses fractional programming to derive precoding coefficients, transmit power, and user-subcarrier association for energy efficiency maximization in a multi-cell system. The paper \cite{large}, instead, investigates the tradeoff between energy efficiency and number of transmit antennas.

This paper considers the downlink of a multi-cell OFDMA system, where a number of BSs form a cluster, share information on channel quality measurements, and collaborate in order to perform energy-efficient user scheduling and power control on the same radio spectrum\footnote{Our approach applies to both frequency- and time-division duplexing.}.
The contributions of this work are summarized as follows.
\begin{itemize}
\item Three figures of merit related to the energy efficiency of the coordinated BSs are considered, namely, the ratio between the sum-rate and the power consumption, which is referred to as global energy-efficiency (GEE), the weighted sum of the energy efficiencies achieved on each resource slot (Sum-EE), and the exponentially-weighted product of the energy efficiencies achieved on each resource slot (Prod-EE). These figures of merit capture different features of the considered communication system, which we illustrate and discuss. Previous related works have mainly focused on the maximization of GEE, but for different system settings. To the best of our knowledge, the work which considers the  scenario most similar to ours is \cite{limited}; however, while \cite{limited} assumes that users are associated to all BSs in the cluster, a configuration usually referred to as virtual (or network) MIMO, we consider a scenario wherein each user is associated to only one BS. As to Sum-EE and Prod-EE, they  have been considered in non-cooperative games \cite{Hierarchy-2009,ZapOFDMA}, but not in the context of coordinated cellular networks.

\item We derive novel procedures aimed at maximizing the above figures of merit with a constraint on the maximum transmit power (either per subcarrier or per base station): this is the major contribution of this work. GEE is optimized by solving a series of concave-convex fractional relaxations, while for Prod-EE a series of concave relaxations is considered. In both cases, the proposed procedures monotonically converge to a solution which at least satisfies the first-order optimality conditions of the original problem. As to Sum-EE, we propose an iterative method to solve the Karush–Kuhn–Tucker (KKT)  conditions of the corresponding non-convex problem. For all figures of merit, we derive algorithms to compute a globally-optimal solution in the asymptotic noise-limited regime.


\item Numerical results indicate that the optimization of the considered figures of merit gives similar performance for low values of the maximum  transmit power; in this case, maximizing the network energy efficiency is also approximatively equivalent to maximizing the network spectral efficiency.  For large values of the maximum transmit power, a moderate reduction of the network spectral efficiency may allow a significant energy saving; in this regime, Sum-EE and Prod-EE  allow to better control the individual energy efficiency achieved by each BS than GEE, which is an attractive feature in heterogeneous networks. Also, Prod-EE ensures a more balanced use of the available subcarriers at the price of a more severe loss in terms of network spectral efficiency.

\end{itemize}

The remainder of this paper is organized as follows. Section~\ref{SEC:Signal model} contains the system description and the problem formulation. Sections~\ref{SEC:GEE}, \ref{SEC:Sum-EE}, and \ref{SEC:Prod-EE} contain the design of the algorithms maximizing GEE, Sum-EE, and Prod-EE, respectively. The numerical results are presented in Section~\ref{SEC:Numerical results}.
 Finally, concluding remarks are given in Section~\ref{SEC:Conclusions}.


\section{System description}\label{SEC:Signal model}
We consider  a cluster of $M$ coordinated BSs in the downlink of an OFDMA network employing $N$  subcarriers and universal frequency reuse. Users and BSs are equipped with one receive and one transmit antenna, respectively.  Each user is connected to only one BS, which is selected based on long-term channel quality measurements.  We denote by ${\cal B}_m$ the (non-empty) set of users assigned to BS $m$ and assume that each BS serves at most one user at a time on each subcarrier. We consider an infinitely backlogged traffic model wherein each access point always has data available for transmission to all connected users. Also, we assume  that the channels remain constant during each transmission frame, and that each user can accurately estimate the channels from the coordinated BSs to itself and feedback them to its serving BS. 

\subsection{Signal model}
Assuming perfect synchronization, the discrete-time baseband signal received
by user $s\in {\cal B}_{m}$ on subcarrier $n$ is
\begin{equation}\label{eq:1}
y_{s}^{[n]}=\underbrace{H_{m,s}^{[n]}x_m^{[n]}}_{
\mbox{\footnotesize in-cell data}}+\underbrace{\sum_{\ell=1,\ell\neq m}^M
H_{\ell,s}^{[n]}x_\ell^{[n]}}_{\mbox{\footnotesize out-of-cell
data}}+\underbrace{n_{s}^{[n]}}_{\mbox{\footnotesize
noise}}\,.\end{equation}
In (\ref{eq:1}), $H_{q,s}^{[n]}$ is the complex channel response between BS $q$ and user $s$ on
subcarrier $n$, which includes small scale fading, large scale fading and path attenuation \cite{Goldsmith_book}, while
$x_q^{[n]}$ is  the complex symbol transmitted by
BS $q$ on subcarrier $n$. The transmitted symbols are modeled as independent random variables with zero mean and variance $\text{E}\{|x_m^{[n]}|^2\}= p_m^{[n]}\geq0$
Finally, $n_{s}^{[n]}$ is the additive noise received by user $s$, which is  modeled as a circularly-symmetric, Gaussian random variable with variance $\mathcal{N}_{s}^{[n]}/2$ per real
dimension. Different noise levels at each
mobile account for different levels of the out-of-cluster interference
and for different noise figures of the receivers.

The signal-to-interference-plus-noise ratio (SINR) for
BS $m$ on subcarrier $n$ when serving user $s\in {\cal B}_{m}$
is \begin{equation}\label{eq:2}
\text{SINR}_{m,s}^{[n]}=   \frac{p_m^{[n]}G_{m,s}^{[n]}}{\displaystyle
1+\sum_{\ell=1,\ell\neq m}^M p_\ell^{[n]}G_{\ell,s}^{[n]}}\end{equation} with
$G_{q,s}^{[n]}=|H_{q,s}^{[n]}|^2/\mathcal{N}_{s}^{[n]}$; also, the corresponding achievable information rate (in bit/s) is \cite{Cover_book}
\begin{equation}\label{eq:3}\text{R}_{m,s}^{[n]}= B\log_2\Big[1+\text{SINR}_{m,s}^{[n]}\Big]\end{equation}
where $B$ is the bandwidth of each subcarrier.

\subsection{Power model}
Following  \cite{circuit_power_consumption,Isheden2011,howmuch2011}, the consumed power is modeled as the sum of two terms, accounting for the power dissipated in the amplifier and in the RF transmit circuits, respectively. The power dissipated in the amplifier is expressed as $\gamma p$, with $p\geq 0$ and $\gamma\geq 1$ being the transmit power output by the amplifier and a scaling coefficient which accounts for the amplifier and feeder losses. Instead, the power dissipated in the remaining circuit blocks is modeled as a constant term $\theta>0$, which accounts for battery backup and for signal processing carried out in the mixer, frequency synthesizer, active filters, and digital-to-analog converter. Both $\theta$ and $\gamma$ generally scale with the additional losses incurred by the power supply and/or the cooling equipment. 
Accordingly,  the power consumed by BS $m$ on subcarrier $n$ is written as
\begin{equation}
P_{c,m}^{[n]}=\theta_{m}^{[n]}+\gamma_{m}^{[n]}p_{m}^{[n]}\;.
\end{equation}

\subsection{Energy-efficient resource allocation}\label{SEC:Energy efficiency}
Let $k(m,n)\in{\cal B}_m$ indicate the user scheduled by BS $m$ on subcarrier $n$ and define $\mathbf{k}^{[n]}=(k(1,n),\ldots,k(M,n))^T$ and $\mathbf{k}= \mbox{vec}\{\mathbf{k}^{[1]}\ldots,\mathbf{k}^{[N]}\}$. Also, we define $\mathbf{p}^{[n]}
=(p_1^{[n]},\ldots,p_M^{[n]})^T$ and $\mathbf{p}=\mbox{vec}\{\mathbf{p}^{[1]},\ldots,\mathbf{p}^{[N]}\}$. System optimization requires selecting $\mathbf{k}$ and $\mathbf{p}$
so to maximize a meaningful figure of merit under some physical constraint.

In this work, we aim at maximizing the network energy efficiency. We consider three figures of merit, which encapsulate different aspects related to the energy efficiency of the considered coordinated cluster. The first one is the global energy efficiency, defined as the ratio between the network sum-rate and the network power consumption, i.e,
\begin{eqnarray}
\text{GEE}(\mathbf{p},\mathbf{k})&=&
\displaystyle \frac{\displaystyle  \sum_{m=1}^{M}\sum_{n=1}^{N}\text{R}_{m, k(m,n)}^{[n]}}{\displaystyle \sum_{m=1}^{M}\sum_{n=1}^N \left(\theta_{m}^{[n]}+\gamma_{m}^{[n]}p_{m}^{[n]}\right)}. \label{eq:GEE}
\end{eqnarray}

Another meaningful figure of merit is the weighted sum of the energy efficiencies across all subcarriers and BSs, i.e.,
\begin{equation}\label{eq:4}
\text{Sum-EE}(\mathbf{p},\mathbf{k})=\sum_{m=1}^{M}\sum_{n=1}^{N}w^{[n]}_{m, k(m,n)} \frac{\text{R}_{m, k(m,n)}^{[n]}}{\theta_{m}^{[n]}+\gamma_{m}^{[n]}p_{m}^{[n]}}
\end{equation} where the weight $w^{[n]}_{m, s}$, for $m=1,\ldots,M$, $n=1,\ldots,N$ and $s\in{\cal B}_m$, may account for the priority of the scheduled users, the nature of the coordinated BSs, and the services assigned to each subcarrier. Differently from GEE, this figure of merit is well suited for heterogenous networks, as the weights can now be used to control the energy efficiency achieved on a specific subcarrier or BS. If we choose $w^{[n]}_{m, s}=\frac{1}{MN}$, then Sum-EE is the arithmetic mean of the energy efficiencies across all subcarriers and BSs.

Finally, we consider the exponentially-weighted product of the energy efficiencies across all subcarriers and BSs, i.e.,
 \begin{equation}\label{eq:prodEE}
\text{Prod-EE}(\mathbf{p},\mathbf{k})=\prod_{m=1}^{M}\prod_{n=1}^{N}\left( \frac{\text{R}_{m, k(m,n)}^{[n]}}{\theta_{m}^{[n]}+\gamma_{m}^{[n]}p_{m}^{[n]}}\right)^{w^{[n]}_{m, k(m,n)}}.
\end{equation}
Due to its multiplicative nature, maximization of (\ref{eq:prodEE}) leads to a configuration where all subcarriers are always used for transmission by all BSs, which may not be the case when GEE or Sum-EE are considered. Therefore, maximizing Prod-EE leads to a more balanced power allocation on the different subcarriers, allowing a simpler design of the transmit amplifiers. Prod-EE also grants the possibility to tune the energy efficiency of each subcarrier through the choice of the weights. For $w^{[n]}_{m, s}=\frac{1}{MN}$, Prod-EE is the geometric mean of the energy efficiencies across all subcarriers and BSs.

In the following, we present algorithms aimed at maximizing the above figures of merit under per-BS or per-subcarrier power constraints. In keeping with a common trend in the open literature, we assume perfect channel state information and optimize the GEE, Sum-EE, and Prod-EE based on instantaneous channels. An alternative approach, that is however out of the scope of this work, is to perform resource allocation based on long-term variations of the channel \cite{conti2007slow,li2010slow}.

The proposed algorithms require to run in a centralized controller, which collects the channel measurements from the coordinated BSs and outputs the scheduling and the transmit power for each BS and subcarrier. This complexity overhead is expected to be affordable with the currently available technology. Indeed, with the advent of the software defined networking paradigm and of the cloud radio access network architecture, cellular systems will be made of light BSs performing only baseband to radio frequency conversion, while neighboring BSs will be connected via high-capacity links to a central unit performing most of the data processing \cite{zhu2011virtual,greensoft}. Clearly, our methods well fit in this context.\footnote{Notice that the Coordinated Multi-Point (CoMP) transmission has been recently introduced in LTE-Advanced \cite{lee2012coordinated}; the CoMP transmission involves a higher degree of cooperation and information sharing among BSs than the proposed resource allocation schemes, yet it is feasible.} Also, BS coordination is usually required only for mobile users that are at the edge of the cells, i.e., midway among two or more BSs; as a consequence, the number of mobile terminals involved may be only a small fraction of the overall set of active users.


\section{Optimization of GEE}\label{SEC:GEE}

In this section, we study the maximization of \eqref{eq:GEE}. We first consider a per-BS power constraint and, then, we specialize the results to the case of a per-subcarrier power constraint. With reference to the more general per-BS power constraint, the noise-limited scenario is also addressed: considering this problem is interesting, since it leads to simpler resource allocation algorithms that can be employed when the intercell interference is weak and, hence, can be neglected.

\subsection{Per-BS power constraint} \label{SEC:GEE-per-BS}

The problem to be solved is
\begin{equation}\label{eq:5A_LB}\left\{
\begin{array}{rl}
\arg&\displaystyle\max_{\mathbf{p},\mathbf{k}}\text{GEE}(\mathbf{p},\mathbf{k})\\
\mbox{s.t.}& \displaystyle \sum_{n=1}^{N} p_{m}^{[n]}\leq P_{m,\max},\;\;\forall\, m\\
&p_{m}^{[n]}\geq0,\;k(m,n)\in{\cal B}_m,\;\;\forall\, m,n
\end{array}\right.
\end{equation}
where $P_{m,\max}$ is the maximum power that can be radiated by BS $m$.
For any feasible $\mathbf{p}$, the optimization over $\mathbf{k}$ is separable across BSs and subcarriers, and the solution is given by
\begin{equation} \label{eq:13-LB}
\hat{k}(m,n)=\arg\max_{s\in {\cal B}_m}\text{R}^{[n]}_{m, s}
\end{equation}
for $m=1,\ldots,M$ and $n=1,\ldots N$.
Next, observe that for any $z\geq0$ and $\bar{z}\geq0$,\footnote{We use the convention that $\log_2(0)=-\infty$ and $0\log_2(0)=0$.}  the following inequality holds \cite{Papandriopoulus-2006}:
\begin{equation}\label{eq:20-A}
\log_2(1+z)\geq\alpha\log_2z+\beta
\end{equation} where $\alpha$ and $\beta$ are defined as
\begin{equation}\label{eq:20-B}
\alpha=\frac{\bar{z}}{1+\bar{z}},\;
\beta=\log_2(1+\bar{z})-\frac{\bar{z}}{1+\bar{z}}\log_2\bar{z}
\end{equation} and the bound is tight for $z=\bar{z}$. As a consequence, for a given feasible user selection $\mathbf{k}$, the following lower bound to the objective function is obtained
\begin{multline}
\text{GEE}(\mathbf{p},\mathbf{k}) \geq h(\mathbf{p},\mathbf{k})=\\
\frac{\displaystyle \overbrace{B\sum_{m=1}^{M}\sum_{n=1}^{N}\left[\alpha_{m}^{[n]}\log_2\left(\text{SINR}_{m, k(m,n)}^{[n]}\right) + \beta_{m}^{[n]}\right]}^{f(\mathbf{p},\mathbf{k})} }{ \displaystyle \underbrace{\sum_{m=1}^{M}\sum_{n=1}^{N}
\left(\theta_{m}^{[n]}+\gamma_{m}^{[n]}p_{m}^{[n]}\right)}_{g(\mathbf{p})}}
\label{eq:LB-Papandriopoulus}
\end{multline}
where $\alpha_m^{[n]}$ and $\beta_m^{[n]}$  are approximation constants computed as in  (\ref{eq:20-B}) for  some $\bar{z}_m^{[n]}\geq0$ to be specified in the following.

Consider now the transformation $q_m^{[n]}=\ln p_m^{[n]}$, and define $\mathbf{q}^{[n]} =(q_1^{[n]},\ldots,q_M^{[n]})^T$, $\mathbf{q}=\mbox{vec}\{\mathbf{q}^{[1]},\ldots,\mathbf{q}^{[N]}\}$, and ${\cal Q}=\{\mathbf{q}\in \mathbb{R}^{MN}: \sum_{n=1}^{N}\exp\{q_{m}^{[n]}\}\leq P_{m,\max},\;\forall\, m\}$. We have the following result, whose proof is reported in Appendix~\ref{Proof-Lemma-1}.
\begin{lemma}\label{Lemma-1}
$f(\exp\{\mathbf{q}\},\mathbf{k})$ is a concave function of $\mathbf{q}$ with $\max_{\mathbf{q}\in {\cal Q}}f(\exp\{\mathbf{q}\},\mathbf{k})\geq 0$. Also, $g(\exp\{\mathbf{q}\})$ is a positive, convex function of $\mathbf{q}$.
\end{lemma}
Leveraging the above Lemma, we proposed to solve (\ref{eq:5A_LB}) by iteratively optimizing the power allocation according to the lower bound in \eqref{eq:LB-Papandriopoulus}, computing the best user selection according to \eqref{eq:13-LB}, and tightening the bound in \eqref{eq:LB-Papandriopoulus}, as summarized in Algorithm~\ref{alg-Papandriopoulus}.
As a consequence of Lemma~\ref{Lemma-1},  the concave-convex fractional problem in \eqref{eq:problem-LB-Papandriopoulus} can be solved using Dinkelbach's procedure \cite{Dinkelbach1967} outlined in Algorithm~\ref{alg-Dinkelbach-per-BS},\footnote{Here, the variable $\epsilon$ denotes the required tolerance, while the indicator FLAG rules the exit from the iterative repeat cycle. Similar considerations apply to Algorithm \ref{alg-Dinkelbach-per-BS-NL}. } while standard  techniques can be used to solve the concave maximization in \eqref{eq:problem-Dinkelbach-per-BS} \cite{Boyd_book}. As to Algorithm~\ref{alg-Papandriopoulus}, we have the following result, whose proof is reported in Appendix~\ref{Proof-Prop-1}.
\begin{proposition}\label{Prop-1}
Algorithm~\ref{alg-Papandriopoulus} monotonically improves the value of GEE at each iteration and converges.
Also, the solution obtained at convergence satisfies the KKT conditions for (\ref{eq:5A_LB}).
\end{proposition}
Notice that the KKT conditions are first-order necessary conditions for any relative maximizer of (\ref{eq:5A_LB}), as the Slater's constraint qualification holds \cite{Bertsekas/99}.

\begin{algorithm}[!t]
\caption{Proposed procedure to solve \eqref{eq:5A_LB}}
\begin{algorithmic}[1]
\label{alg-Papandriopoulus}
\STATE Initialize $I_{\max}$ and set  $i=0$

\STATE Initialize $\mathbf{p}$ and compute  $\mathbf{k}$  according to (\ref{eq:13-LB})

\REPEAT

\STATE Set $\bar{z}_m^{[n]}=\text{SINR}_{m, k(m,n)}^{[n]}$ and
compute $\alpha_m^{[n]}$ and $\beta_m^{[n]}$ as in  (\ref{eq:20-B}), for $m=1,\ldots,M$ and $n=1,\ldots,N$

\STATE  Update $\mathbf{p}$ by solving the following problem using Algorithm \ref{alg-Dinkelbach-per-BS} ($\mathbf{p}=\exp\{\mathbf{q}\}$):
\begin{equation}\label{eq:problem-LB-Papandriopoulus}
\arg \displaystyle \max_{\mathbf{q}\in {\cal Q}}h(\exp\{\mathbf{q}\},\mathbf{k})
\end{equation}

\STATE Update  $\mathbf{k}$  according to (\ref{eq:13-LB})

\STATE Set  $i=i+1$

\UNTIL{convergence or $i=I_{\max}$}

\end{algorithmic}
\end{algorithm}

\begin{algorithm}[!t]
\caption{Dinkelbach's procedure  \cite{Dinkelbach1967} to solve \eqref{eq:problem-LB-Papandriopoulus}}
\begin{algorithmic}[1]
\label{alg-Dinkelbach-per-BS}
\STATE Set $\epsilon>0$, $\pi=0$, and $\textrm{FLAG}=0$

\REPEAT

\STATE  Update $\mathbf{q}$ by solving the following concave maximization:
\begin{equation}\label{eq:problem-Dinkelbach-per-BS}
\arg \displaystyle \max_{\mathbf{q}\in {\cal Q}} f(\exp\{\mathbf{q}\},\mathbf{k})-\pi g(\exp\{\mathbf{q}\})
\end{equation}

\IF{ $ f(\exp\{\mathbf{q}\},\mathbf{k})-\pi g(\exp\{\mathbf{q}\})< \epsilon $}
  \STATE  $\textrm{FLAG}=1$
\ELSE
\STATE Set $\pi=f(\exp\{\mathbf{q}\},\mathbf{k})/g(\exp\{\mathbf{q}\})$
\ENDIF
\UNTIL{$\textrm{FLAG}=1$}
\end{algorithmic}
\end{algorithm}

\subsubsection{Noise-limited (NL) regime}
Neglecting the intercell interference, GEE simplifies to
\begin{multline}\nonumber
\text{GEE-NL}(\mathbf{p},\mathbf{k})= \displaystyle\dfrac{\overbrace{\displaystyle B \sum_{m=1}^{M}\sum_{n=1}^{N} \log_2\left(1+p_{m}^{[n]}G_{m,k(m,n)}^{[n]}\right)}^{f_{\text{NL}}(\mathbf{p},\mathbf{k})}}
{\displaystyle  \sum_{m=1}^{M}\sum_{n=1}^{N}\left(\theta_{m}^{[n]}+\gamma_{m}^{[n]}p_{m}^{[n]}\right)}
\end{multline}
and the problem to be solved becomes
\begin{equation}\label{eq:GEE-NL-problem}\left\{
\begin{array}{rl}
\arg&\displaystyle\max_{\mathbf{p},\mathbf{k}}\text{GEE-NL}(\mathbf{p},\mathbf{k})\\
\mbox{s.t.}& \displaystyle \sum_{n=1}^{N} p_{m}^{[n]}\leq P_{m,\max},\;\;\forall\, m\\
&p_{m}^{[n]}\geq0,\;k(m,n)\in{\cal B}_m,\;\;\forall\, m,n.
\end{array}\right.
\end{equation}
We now have the following results, whose proof is reported in Appendix~\ref{Proof-Prop-1-NL}.
\begin{proposition}\label{Prop-1-NL}
Algorithm~\ref{alg-Papandriopoulus-NL} monotonically improves the value of $\text{GEE-NL}$ at each iteration, and provides a globally optimal solution to \eqref{eq:GEE-NL-problem}.
\end{proposition}

Notice that the concave-linear fractional problem (\ref{eq:problem-LB-Papandriopoulus-NL}) in Algorithm~\ref{alg-Papandriopoulus-NL} can be solved by using Algorithm~\ref{alg-Dinkelbach-per-BS-NL}. Also, the solution to the concave maximization (\ref{eq:problem-Dinkelbach-per-BS-NL}) can be found from the KKT conditions; in particular, after standard manipulation we obtain the following $M$ waterfilling-like problems (one for each BS)
\begin{equation*}
\left\{
\begin{array}{l}
p_{m}^{[n]}=\max\left\{0, \dfrac{(B/\ln2)}{\pi \gamma^{[n]}_{m, k(m,n)}+\lambda_{m}}-\dfrac{1}{G_{m,k(m,n)}^{[n]}}\right\}\\
\displaystyle\sum_{n=1}^{N}p_{m}^{[n]}\leq P_{m,\max}
\end{array}
\right.
\end{equation*}
for $m=1,\ldots,M$,  and the optimal value of the non-negative Lagrange multiplier $\lambda_{m}$ can be derived by bisection search.

\begin{algorithm}[!t]
\caption{Proposed procedure to solve \eqref{eq:GEE-NL-problem} in the noise limited regime}
\begin{algorithmic}[1]
\label{alg-Papandriopoulus-NL}
\STATE Initialize $I_{\max}$ and set  $i=0$

\STATE Initialize $\mathbf{p}$ and compute  $\mathbf{k}$  according to
\begin{equation}\label{eq:sceduling-NL}
k(m,n)=\arg\max_{s\in {\cal B}_m} \log_2\left(1+p_{m}^{[n]}G_{m,s}^{[n]}\right)
\end{equation}
for $m=1,\ldots,M$ and $n=1,\ldots,N$.
\REPEAT

\STATE  Update $\mathbf{p}$ by solving the following problem using Algorithm~\ref{alg-Dinkelbach-per-BS-NL}:
\begin{equation}\label{eq:problem-LB-Papandriopoulus-NL}\left\{
\begin{array}{rl}
\arg& \displaystyle \max_{\mathbf{p}}\text{GEE-NL}(\mathbf{p},\mathbf{k})\\
\mbox{s.t.}& \sum_{n=1}^{N}p_{m}^{[n]}\leq P_{m,\max},\;\;\forall\, m\\
&p_{m}^{[n]}\geq0,\;\;\forall\, m,n
\end{array}\right.
\end{equation}

\STATE Update  $\mathbf{k}$  according to \eqref{eq:sceduling-NL}

\STATE Set  $i=i+1$

\UNTIL{convergence or $i=I_{\max}$}

\end{algorithmic}
\end{algorithm}

\begin{algorithm}[!t]
\caption{Dinkelbach's procedure \cite{Dinkelbach1967} to solve \eqref{eq:problem-LB-Papandriopoulus-NL}}
\begin{algorithmic}[1]
\label{alg-Dinkelbach-per-BS-NL}
\STATE Set $\epsilon>0$, $\pi=0$, and $\textrm{FLAG}=0$

\REPEAT

\STATE  Update $\mathbf{p}$ by solving the following concave maximization:
\begin{equation}\label{eq:problem-Dinkelbach-per-BS-NL}\left\{
\begin{array}{rl}
\arg& \displaystyle \max_{\mathbf{p}} f_{\text{NL}}(\mathbf{p},\mathbf{k})-\pi g(\mathbf{p})\\
\mbox{s.t.}& \sum_{n=1}^{N}p_{m}^{[n]}\leq P_{m,\max},\;\;\forall\, m\\
&p_{m}^{[n]}\geq0,\;\;\forall\, m,n
\end{array}\right.
\end{equation}
\IF{ $  \displaystyle f_{\text{NL}}(\mathbf{p},\mathbf{k})-\pi g(\mathbf{p}) < \epsilon $}
  \STATE  $\textrm{FLAG}=1$
\ELSE
\STATE Set $  \pi=f_{\text{NL}}(\mathbf{p},\mathbf{k})/g(\mathbf{p})$
\ENDIF
\UNTIL{$\textrm{FLAG}=1$}
\end{algorithmic}
\end{algorithm}

\subsection{Per-subcarrier power constraint}
The problem to be solved is
\begin{equation*}\left\{
\begin{array}{rl}
\arg&\displaystyle\max_{\mathbf{p},\mathbf{k}}\text{GEE}(\mathbf{p},\mathbf{k})\\
\mbox{s.t.}&\displaystyle 0\leq p_{m}^{[n]}\leq P_{m,\max}^{[n]},\;k(m,n)\in{\cal B}_m,\;\;\forall\, m,n.
\end{array}\right.
\end{equation*}
where $P_{m,\max}^{[n]}$ is the maximum power that can be radiated by BS $m$ on subcarrier $n$.
GEE is not a separable function of $\mathbf{p}$, whereby the above optimization problem does not decouple across subcarriers.
Luckily enough, all derivations carried out in Section~\ref{SEC:GEE-per-BS} can be replicated here with minor modifications.
In particular, Algorithms~\ref{alg-Papandriopoulus} and \ref{alg-Dinkelbach-per-BS} remain the same, except that the feasible set ${\cal Q}$ in (\ref{eq:problem-LB-Papandriopoulus}) and (\ref{eq:problem-Dinkelbach-per-BS}) must be re-defined as ${\cal Q}=\{\mathbf{q}\in \mathbb{R}^{MN}: \exp\{q_{m}^{[n]}\}\leq P_{m,\max}^{[n]},\;\forall\, m,n\}$. Interestingly, the  solution to the concave maximization in (\ref{eq:problem-Dinkelbach-per-BS}) can now be computed with a simple iterative method. Indeed, consider the following KKT conditions for (\ref{eq:problem-Dinkelbach-per-BS}) \cite{Boyd_book}:
\begin{align}
&\displaystyle
\begin{array}{rr}
\displaystyle \frac{\text{d}}{\displaystyle \text{d} q_m^{[n]}}\left(f(\exp\{\mathbf{q}\},\mathbf{k})-\pi g(\exp\{\mathbf{q}\})\right)-\\\lambda_{m}^{[n]}\exp\{q_{m}^{[n]}\}=0,
\;\forall\;m,n
\end{array}\label{eq:KKT-problem-Dinkelbach-per-tone}
\\
&\lambda_{m}^{[n]}\geq0,\;\forall\;m,n \nonumber \\
&\exp\{q_{m}^{[n]}\}\leq P_{m,\max}^{[n]},\;\forall\;m,n \nonumber \\
&\lambda_{m}^{[n]}\left(P_{m,\max}^{[n]}-\exp\{q_{m}^{[n]}\}\right)=0,\;\forall\;m,n \nonumber
\end{align}
where $\lambda_{m}^{[n]}$ is the Lagrange multiplier associated to the power constraint of BS $m$ on subcarrier $n$. After some manipulations,  the stationary condition in (\ref{eq:KKT-problem-Dinkelbach-per-tone}) can be recast as
\begin{multline}\label{eq:21}
\exp\{q_{m}^{[n]}\}= \alpha_{m}^{[n]}B \Bigg[\left(\gamma_{m}^{[n]}\pi+\lambda_{m}^{[n]} \right)\ln2+ \\
B \sum_{j=1,j\neq m}^{M}
\frac{ \displaystyle \alpha_{j}^{[n]}G_{m,k(j,n)}^{[n]} }{ 1+\sum_{\ell=1,\ell\neq j}^{M} \exp\{q_{\ell}^{[n]}\}G_{\ell,k(j,n)}^{[n]}} \Bigg]^{-1}.
\end{multline}
Since the right hand side  (RHS) of (\ref{eq:21}) is a standard interference function \cite{Yates-1995}, the optimal $\mathbf{q}$ can be obtained by starting from any feasible power allocation and iteratively solving the following fixed point equations:
\begin{multline}\nonumber
\exp\{q_{m}^{[n]}\}=\min\left\{\rule{0mm}{1.2cm}P_{m,\max}^{[n]}, \right.
\\ \left.
\frac{\displaystyle \alpha_{m}^{[n]}B}{\displaystyle
\gamma_{m}^{[n]}\pi\ln2\!+\! B \!\! \!\sum_{j=1,j\neq m}^{M} \!\!
\frac{ \displaystyle \alpha_{j}^{[n]}G_{m,k(j,n)}^{[n]} }{ 1+\sum_{\ell=1,\ell\neq j}^{M}\exp\{q_{\ell}^{[n]}\}G_{\ell,k(j,n)}^{[n]}}
}\right\}
\end{multline}
for $m=1,\ldots,M$ and $n=1,\ldots,N$.

\section{Optimization of Sum-EE}\label{SEC:Sum-EE}

In this section, we study the maximization of \eqref{eq:4} under a per-BS power constraint, i.e.,
\begin{equation}\label{eq:5B}\left\{
\begin{array}{rl}
\arg& \displaystyle \max_{\mathbf{p},\mathbf{k}} \text{Sum-EE}(\mathbf{p},\mathbf{k})\\
\mbox{s.t.}& \displaystyle\sum_{n=1}^N p_{m}^{[n]}\leq P_{m,\max},\;\;\forall\, m\\
&p_{m}^{[n]}\geq0,\;k(m,n)\in{\cal B}_m,\;\;\forall\, m,n.
\end{array}\right.
\end{equation}
Since Sum-EE is a separable function of the power variables, the following results specialize to a per-subcarrier power constraint in a straightforward manner; the corresponding details are omitted for brevity. Paralleling Section \ref{SEC:GEE}, Problem (\ref{eq:5B}) is also studied in the noise-limited scenario.

For a given feasible $\mathbf{p}$, the optimization over $\mathbf{k}$ is separable across BSs and subcarriers, and the solution is given by
\begin{equation}
\label{eq:user-update-SUM-EE}
\hat{k}(m,n)=\arg\max_{s\in {\cal B}_m}w_{m,s}^{[n]}\text{R}^{[n]}_{m, s}
\end{equation}
for $m=1,\ldots,M$ and $n=1,\ldots N$. On the other hand, for any feasible user selection, the optimal set of powers must  satisfy the following KKT conditions:

\begin{subequations}\label{eq:7B}
\begin{align}\displaystyle
&\frac{\text{d}}{\displaystyle \text{d} p_m^{[n]}}\text{Sum-EE}(\mathbf{p},\mathbf{k})
+\mu_{m}^{[n]}-\lambda_{m}=0,\;\forall\;m,n \label{eq:7B-stazionary}\\
&\mu_{m}^{[n]}\geq0,\;\; \lambda_{m}\geq0,\;\forall\;m,n \label{eq:7B-multi}\\
&-p_m^{[n]}\leq0,\;\forall\;m,n \label{eq:7B-power-1}\\
&\sum_{n=1}^N p_m^{[n]}\leq P_{m,\max} ,\;\forall\;m \label{eq:7B-power-2}\\
&\mu_{m}^{[n]}p_m^{[n]}=0,\;\forall\;m,n \label{eq:7B-compl-1}\\
&\lambda_{m}\left(P_{m,\max}-\sum_{n=1}^N p_m^{[n]}\right)=0,\;\forall\;m \label{eq:7B-compl-2}
\end{align}
\end{subequations}
where $\lambda_{m}$ and $\mu_{m}^{[n]}$ are the Lagrange multipliers associated to the constraints on the maximum power radiated by BS $m$ and on the minimum power level of BS $m$ on subcarrier $n$, respectively.

After standard algebraic manipulations, it can be shown that
\begin{multline}\nonumber
\displaystyle
\frac{\text{d}}{\displaystyle \text{d} p_m^{[n]}}\text{Sum-EE}(\mathbf{p},\mathbf{k})=(B/\ln2)\text{Q}_{m,k(m,n)}^{[n]}\times \\
\displaystyle \frac{G_{m,k(m,n)}^{[n]}}{\displaystyle 1+\text{I}_{m,k(m,n)}^{[n]}+p_m^{[n]}G_{m,k(m,n)}^{[n]}}-\text{C}_{m,k(m,n)}^{[n]}-\text{L}_{m,k(m,n)}^{[n]}
\end{multline}
where
\begin{align}
\label{eq:I}
\text{I}_{m,k(m,n)}^{[n]}&=\sum_{\ell=1,\ell\neq m}^{M}\!\!\!p_\ell^{[n]}G_{\ell,k(m,n)}^{[n]}\\
\label{eq:Q}
\text{Q}_{m,k(m,n)}^{[n]}&=\dfrac{w_{m,k(m,n)}^{[n]}}{\theta_{m}^{[n]}+\gamma^{[n]}_{m}p_{m}^{[n]}}\\
\label{eq:Z}
\text{C}_{m,k(m,n)}^{[n]}&= w^{[n]}_{m, k(m,n)}\gamma^{[n]}_{m}
\dfrac{\text{R}_{m,k(m,n)}^{[n]}}{\left(\theta_{m}^{[n]}+\gamma^{[n]}_{m}p_{m}^{[n]}\right)^2}\\
\label{eq:T}
\text{L}_{m,k(m,n)}^{[n]}&=\frac{B}{\ln2}\sum_{j=1,j\neq m}^{M}\text{Q}_{j, k(j,n)}^{[n]}
\frac{ G_{m,k(j,n)}^{[n]} \text{SINR}_{j, k(j,n)}^{[n]} }{\displaystyle
1+\sum_{\ell=1}^M p_\ell^{[n]}G_{\ell,k(j,n)}^{[n]}}
\end{align}
whereby \eqref{eq:7B-stazionary} can be rewritten as
\begin{equation}\label{eq:12}
p_{m}^{[n]}=\displaystyle\frac{(B/\ln2) Q^{[n]}_{m,k(m,n)}}{\lambda_{m}-\mu_{m}^{[n]}+\text{C}_{m,k(m,n)}^{[n]}+\text{L}_{m,k(m,n)}^{[n]}}
-\displaystyle\frac{\displaystyle  1+\text{I}_{m,k(m,n)}^{[n]}}{G_{m,k(m,n)}^{[n]}}.
\end{equation}
Notice that $\text{I}_{m,k(m,n)}^{[n]}$ is the co-channel interference for the user scheduled by BS $m$ on sub-carrier $n$;  $\text{Q}_{m,k(m,n)}^{[n]}$ is an equivalent weight for the rate achieved by BS $m$ on sub-carrier $n$, scaled by the corresponding power consumption; $\text{C}_{m,k(m,n)}^{[n]}$ is a marginal power cost paid by BS $m$ for transmitting on subcarrier $n$ to user $k(m,n)$; finally, $\text{L}_{m,k(m,n)}^{[n]}$ accounts for the interference leakage to undesired receivers when serving user $k(m,n)$. The stationary condition \eqref{eq:12} indicates that, for any given set of scheduled users $\mathbf{k}$, BS $m$ should allocate more power to subcarriers having larger equivalent weights and experiencing better channel conditions; also, the taxation terms $\text{C}_{m,k(m,n)}^{[n]}$ and $\text{L}_{m,k(m,n)}^{[n]}$ lower the radiated power if the marginal power price paid by BS $m$ to transmit on subcarrier $n$ to user $k(m,n)$ is large and if this transmission causes an excessive leakage to other co-channel users, respectively.

Inspired by the modified waterfilling methods considered in \cite{Yu-2007,venturino-2009} for the weighted sum-rate maximization, we now provide an iterative procedure to solve (\ref{eq:user-update-SUM-EE}) and (\ref{eq:7B}), which together are the first-order necessary conditions for the optimal solutions to (\ref{eq:5B}), as the Slater's constraint qualification holds \cite{Bertsekas/99}. Assume that some feasible $\mathbf{p}$ and $\mathbf{k}$ are given. Then, the equivalent weight $\text{Q}_{m,k(m,n)}^{[n]}$, the marginal power cost $\text{C}_{m,k(m,n)}^{[n]}$, and the interference leakage $\text{L}_{m,k(m,n)}^{[n]}$ can be computed from \eqref{eq:Q},  \eqref{eq:Z}, and
\eqref{eq:T}, respectively, for $m=1,\ldots,M$ and $n=1,\ldots,N$. Then, each BS can compute the co-channel interference levels on each subcarrier from \eqref{eq:I} and update the corresponding radiated powers  using \eqref{eq:12}; notice that the Lagrange multipliers must be chosen so as to satisfy the power constraints \eqref{eq:7B-power-1}-\eqref{eq:7B-power-2} and the corresponding complementary slackness conditions \eqref{eq:7B-compl-1}-\eqref{eq:7B-compl-2}, resulting in the following waterfilling-like problems (one for each BSs):
\begin{equation}\label{eq:modified-WF-2}
\left\{
\begin{array}{l}
\begin{array}{rr}
p_{m}^{[n]}=
\max\left\{0,\displaystyle \frac{(B/\ln2)\text{Q}_{m,k(m,n)}^{[n]}}{\lambda_{m}+\text{C}_{m,k(m,n)}^{[n]}+\text{L}_{m,k(m,n)}^{[n]}}
- \right. \\ \left.
\displaystyle\frac{\displaystyle  1+\text{I}_{m,k(m,n)}^{[n]}}{G_{m,k(m,n)}^{[n]}}\right\}
\end{array}
\\
\displaystyle \sum_{n=1}^N p_m^{[n]}\leq P_{m,\max}
\end{array}
\right.
\end{equation}
for $m=1,\ldots,M$, which in turn can be efficiently solved by bisection search. After updating the power level, the scheduled users can be recomputed according to \eqref{eq:user-update-SUM-EE}, and the entire process can be iterated as summarized in Algorithm~\ref{alg-2}

Deriving general conditions under which Algorithm~\ref{alg-2} provably converges seems intractable. Nevertheless, should Algorithm~\ref{alg-2} converge to a feasible solution, then the corresponding set of radiated powers, scheduled users, and Lagrange multipliers satisfy by construction the KKT conditions in (\ref{eq:user-update-SUM-EE}) and (\ref{eq:7B}). Algorithm~\ref{alg-2} can be modified to enforce monotonic convergence by performing the power update at line \ref{power-update-2} only if the objective function is non decreased.\footnote{In our experiments in Section \ref{SEC:Numerical results} we have always observed convergence of Algorithm~\ref{alg-2} without performing this modification.} However, in this latter case, the solution at convergence is not guaranteed to  simultaneously satisfy (\ref{eq:user-update-SUM-EE}) and (\ref{eq:7B}).

\begin{algorithm}[!t]
\caption{Proposed procedure to solve \eqref{eq:5B}}
\label{alg-2}
\begin{algorithmic}[1]

\STATE Initialize $I_{\max}$ and set  $i=0$

\STATE Initialize $\mathbf{p}$ and compute $\mathbf{k}$  according to
\eqref{eq:user-update-SUM-EE}

\REPEAT

\STATE Compute  $\text{Q}_{m,k(m,n)}^{[n]}$, $\text{C}_{m,k(m,n)}^{[n]}$ and $\text{L}_{m,k(m,n)}^{[n]}$, for $m=1,\ldots,M$ and $n=1,\ldots,N$, according to (\ref{eq:Q}), (\ref{eq:Z}) and (\ref{eq:T}), respectively
\FOR{$m=1$ to $M$}
\STATE Compute  $\text{I}_{m,k(m,1)}^{[1]},\ldots, \text{I}_{m,k(m,N)}^{[N]}$ according to (\ref{eq:I})
\STATE \label{power-update-2} Update $p_m^{[1]},\ldots,p_m^{[N]}$ according to \eqref{eq:modified-WF-2}
\ENDFOR
\STATE Update $\mathbf{k}$  according to (\ref{eq:user-update-SUM-EE})

\STATE $i=i+1$
\UNTIL{ convergence or $i=I_{\max}$}
\end{algorithmic}
\end{algorithm}

\subsection{Insights into Algorithm~\ref{alg-2}}

Let $\text{I}_{m,k(m,n)}^{[n]}$, $\text{Q}_{m,k(m,n)}^{[n]}$, $\text{C}_{m,k(m,n)}^{[n]}$, and  $\text{L}_{m,k(m,n)}^{[n]}$
be preassigned and fixed, for $m=1,\ldots,M$ and $n=1,\ldots,N$.
For a given set of scheduled users $\mathbf{k}$, a set of power levels which satisfy \eqref{eq:7B-multi}-\eqref{eq:7B-compl-2} and  \eqref{eq:12}
must also satisfy the KKT conditions of the following problems (one for each BS):
\begin{equation}\label{eq:insight-problem}\left\{
\begin{array}{rl}
\arg& \displaystyle \max_{p_m^{[1]},\ldots,p_m^{[N]}} \sum_{n=1}^{N}\text{Q}_{m,k(m,n)}^{[n]}B \log_2\left(1+\frac{p_{m}^{[n]}G_{m,k(m,n)}^{[n]}}{1+\text{I}_{m,k(m,n)}^{[n]}}\right)-\\ &\displaystyle\hspace{1.5cm}\sum_{n=1}^{N}\left[\text{C}_{m,k(m,n)}^{[n]}+\text{L}_{m,k(m,n)}^{[n]}\right]p_{m}^{[n]}\\
\mbox{s.t.}& \displaystyle\sum_{n=1}^N p_{m}^{[n]}\leq P_{m,\max},\; p_{m}^{[n]}\geq0
\end{array}\right.
\end{equation}
for $m=1,\ldots,M$. The first term of the objective function in \eqref{eq:insight-problem} is a weighted sum of the rates achieved by BS $m$ on its subcarriers with weights $\text{Q}_{m,k(m,n)}^{[n]}$ and interference levels $\text{I}_{m,k(m,n)}^{[n]}$. Also, $\text{C}_{m,k(m,n)}^{[n]}p_{m}^{[n]}$ and $\text{L}_{m,k(m,n)}^{[n]}p_{m}^{[n]}$ are
costs paid by BS $m$ to serve user $k(m,n)$ scheduled on subcarrier $n$ due to the power consumption (i.e., the energy efficiency) and the interference caused to other co-channel users, respectively. Notice that \eqref{eq:insight-problem} is a concave maximization; consequently, the Lagrange multiplier $\lambda_m$ and the power levels $p_m^{[1]},\ldots,p_m^{[N]}$ computed at line \ref{power-update-2} of Algorithm~\ref{alg-2} can also be found by solving \eqref{eq:insight-problem} with any convex optimization tool \cite{Boyd_book}.

\subsection{Noise-limited regime} \label{SEC:Sum-EE-NL}
Neglecting the intercell interference, Sum-EE simplifies to
\begin{multline}\label{eq:Sum-EE-NL}
\text{Sum-EE-NL}(\mathbf{p},\mathbf{k})=\\
\displaystyle B\sum_{m=1}^{M}\sum_{n=1}^{N}w^{[n]}_{m, k(m,n)} \frac{\log_{2}\left(1+p_{m}^{[n]}G_{m,k(m,n)}^{[n]}\right)}{\theta_{m}^{[n]}+\gamma_{m}^{[n]}p_{m}^{[n]}}
\end{multline}
which is a separable function with respect to both $\mathbf{p}$ and $\mathbf{k}$.
For a fixed $\mathbf{p}$, a solution to
\begin{equation*}\left\{
\begin{array}{rl}
\arg& \displaystyle \max_{\mathbf{k}} \text{Sum-EE-NL}(\mathbf{p},\mathbf{k}) \\
\mbox{s.t.}& k(m,n)\in{\cal B}_m,\;\;\forall\, m,n.
\end{array}\right.
\end{equation*}
is given by
\begin{equation}\label{eq:sceduling-SUM-EE-NL}
k(m,n)=\arg\max_{s\in {\cal B}_m} w_{m,s}^{[n]}\log_2\left(1+p_{m}^{[n]}G_{m,s}^{[n]}\right)
\end{equation}
for $m=1,\ldots,M$ and $n=1,\ldots,N$. Also, for a fixed $\mathbf{k}$, the relaxed problem
\begin{equation}\label{Prob:NoiseLimitedRelaxed}\left\{
\begin{array}{rl}
\arg& \displaystyle \max_{\mathbf{p}} \text{Sum-EE-NL}(\mathbf{p},\mathbf{k}) \\
\mbox{s.t.}& p_{m}^{[n]}\geq0,\;\;\forall\, m,n
\end{array}\right.
\end{equation}
has a unique solution, say $\bar{\mathbf{p}}$, which is found by separately maximizing each summand in the objective function
 \cite{Avriel-book}.
We now give the following result, whose proof is reported in Appendix~\ref{Proof-Lemma-2}.
\begin{lemma}\label{Prop:RegionConcave}
Let $u(x)=\displaystyle\frac{\log_{2}(1+ax)}{x+c}$ with $a$ and $c$ positive constants. The function  $u$
is concave in the region $0\leq x \leq \bar{x}$, where $\bar{x}$ is the unique solution to $\arg\max_{x\geq0}u(x)$.
\end{lemma}

Using Lemma \ref{Prop:RegionConcave}, the power allocation problem in the noise-limited regime can be recast as a concave problem
\begin{equation}\label{Prob:NoiseLimitedEE2}
\left\{
\begin{array}{rl}
\arg& \displaystyle \max_{\mathbf{p}} \text{Sum-EE-NL}(\mathbf{p},\mathbf{k})\\
\mbox{s.t.}& \sum_{n=1}^{N}p_{m}^{[n]}\leq P_{m,\max},\;\;\forall m\\
&0 \leq p_{m}^{[n]}\leq \bar{p}_{m}^{[n]},\;\;\forall\, m,n\\
\end{array}\right.
\end{equation}
where $\bar{\mathbf{p}}$ is the solution to \eqref{Prob:NoiseLimitedRelaxed}. Consequently, the optimal resource allocation can be found by alternate maximization of \eqref{eq:Sum-EE-NL} over the variables $\mathbf{k}$ and $\mathbf{p}$ according to \eqref{eq:sceduling-SUM-EE-NL} and \eqref{Prob:NoiseLimitedEE2}, respectively.

\section{Optimization of Prod-EE}\label{SEC:Prod-EE}

In this section, we study the maximization of \eqref{eq:prodEE} under a per-BS power constraint. Since Prod-EE is separable with respect to $\mathbf{p}$, the following results can be specialized to a per-subcarrier power constraint in a straightforward manner.

The problem to be solved is
\begin{equation}\label{Prob:ProdLog2}\left\{
\begin{array}{rl}
\arg& \displaystyle \max_{\mathbf{p},\mathbf{k}}  \ln \text{Prod-EE}(\mathbf{p},\mathbf{k})
\\
\mbox{s.t.}& \displaystyle\sum_{n=1}^N p_{m}^{[n]}\leq P_{m,\max},\;\;\forall\, m\\
&p_{m}^{[n]}\geq0,\;\; k(m,n)\in{\cal B}_m,\;\;\forall\, m,n
\end{array}\right.
\end{equation}
where, without loss of optimality, the objective function is the logarithm of  \eqref{eq:prodEE}. Notice first that a solution to \eqref{Prob:ProdLog2} must necessarily have $\text{R}_{m,k(m,n)}^{[n]}>0$ for $m=1,\ldots,M$ and $n=1,\ldots,N$. Also, for any feasible $\mathbf{p}$, the maximization with respect to $\mathbf{k}$ decouples across BSs and subcarriers, yielding
\begin{equation}\label{Eq:ProdTones}
k(m,n)=\arg\max_{s\in {\cal B}_m}
w^{[n]}_{m, s}\ln\dfrac{\text{R}_{m,s}^{[n]}}{\theta_{m}^{[n]}+\gamma_{m}^{[n]}p_{m}^{[n]}}
\end{equation}
for $m=1,\ldots,M$ and $n=1,\ldots,N$. As to the optimization with respect to $\mathbf{p}$, we consider the following lower bound to the objective function
\begin{multline}\label{eq:Prod-EE-Bound}
\ln \text{Prod-EE}(\mathbf{p},\mathbf{k})\geq\phi(\mathbf{p},\mathbf{k})=
\displaystyle \sum_{m=1}^{M}\sum_{n=1}^{N}w^{[n]}_{m, k(m,n)} \times \\ \displaystyle \ln\left(\frac{\alpha_{m}^{[n]}\log_2\left(\text{SINR}_{m,k(m,n)}^{[n]}\right) + \beta_{m}^{[n]}}{\left(\theta_{m}^{[n]}+\gamma_{m}^{[n]}p_{m}^{[n]}\right)/B}\right)
\end{multline} where $\alpha_m^{[n]}$ and $\beta_m^{[n]}$ are computed as in  (\ref{eq:20-B}) for  some
$\bar{z}_m^{[n]}\geq0$ to be specified in the following. The above bound holds for all $\mathbf{p}$ such that the argument of $\ln$ is non-negative.
Using the transformation $\mathbf{p}=\exp\{\mathbf{q}\}$, the following relaxed power allocation problem is obtained:
\begin{equation}\label{Prob:ProdLog3}\left\{
\begin{array}{lll}
\arg& \displaystyle \max_{\mathbf{q}}  \phi(\exp\{\mathbf{q}\},\mathbf{k}) \\
\mbox{s.t.}& \displaystyle\sum_{n=1}^N e^{q_{m}^{[n]}}\leq P_{m,\max},\;\;\forall\, m\\
&
\begin{array}{rr}
\alpha_{m}^{[n]}
\displaystyle \log_2\left(\frac{G_{m,k(m,n)}^{[n]}\exp\{q_{m}^{[n]}\}}
{1\!\!+\!\!\!\displaystyle\sum_{\ell=1,\,\ell\neq m}^{M}\!\!\! G_{\ell,k(m,n)}^{[n]}\exp\{q_{\ell}^{[n]}\}}\right)+\beta_{m}^{[n]}\geq 0 ,\\ \forall\, m,n.
\end{array}
\end{array}\right.
\end{equation}
Since log-sum-exp is convex, the constrained maximization in \eqref{Prob:ProdLog3} is concave and, hence, can be solved using standard techniques \cite{Boyd_book}.
\algsetup{indent=1.5em}
\begin{algorithm}[t]
\caption{Proposed procedure to solve \eqref{Prob:ProdLog2}}
\begin{algorithmic}[1]
\label{alg-ProdLog1}
\STATE Initialize $I_{\max}$ and set  $i=0$
\STATE Initialize $\mathbf{p}$ and compute  $\mathbf{k}$  according to \eqref{Eq:ProdTones}
\REPEAT
\STATE Set $\bar{z}_m^{[n]}=\text{SINR}_{m, k(m,n)}^{[n]}$ and
compute $\alpha_m^{[n]}$ and $\beta_m^{[n]}$ as in  (\ref{eq:20-B}), for $m=1,\ldots,M$ and $n=1,\ldots,N$
\STATE  Compute $\mathbf{q}$ as the solution to the concave problem (\ref{Prob:ProdLog3}) and update $\displaystyle\mathbf{p}=\exp\{\mathbf{q}\}$
\STATE Update  $\mathbf{k}$  according to \eqref{Eq:ProdTones}
\STATE Set  $i=i+1$
\UNTIL{convergence or $i=I_{\max}$}
\end{algorithmic}
\end{algorithm}

We now propose to solve (\ref{Prob:ProdLog2}) by iteratively optimizing the power allocation according to \eqref{Prob:ProdLog3}, computing the best user selection according to \eqref{Eq:ProdTones}, and tightening the bound in \eqref{eq:Prod-EE-Bound}, as summarized in Algorithm~\ref{alg-ProdLog1}. The following convergence result now holds; the proof is similar to that of Proposition \ref{Prop-1} and is omitted for brevity.
\begin{proposition}
Algorithm~\ref{alg-ProdLog1} monotonically improves the value of Prod-EE at each iteration and converges.
Also, the solution obtained at convergence satisfies the KKT conditions for (\ref{Prob:ProdLog2}).
\end{proposition}
Notice again that, since the Slater's constraint qualification holds \cite{Bertsekas/99}, the KKT conditions are first-order necessary conditions for any relative maximizer of (\ref{Prob:ProdLog2}).

As to the noise limited regime, notice that the objective function in \eqref{Prob:ProdLog2} simplifies to
\begin{equation*}\label{eq:Prod-EE-NL}
\sum_{m=1}^{M}\sum_{n=1}^{N}w^{[n]}_{m, k(m,n)}
\ln\left(\frac{B\log_{2}\left(1+p_{m}^{[n]}G_{m,k(m,n)}^{[n]}\right)}{\theta_{m}^{[n]}+\gamma_{m}^{[n]}p_{m}^{[n]}} \right)
\end{equation*} and all derivations in Section \ref{SEC:Sum-EE-NL} can be replicated here.

\section{Numerical results}\label{SEC:Numerical results}

\begin{figure}[!tbp]
\centerline{\includegraphics[width=0.95\columnwidth]{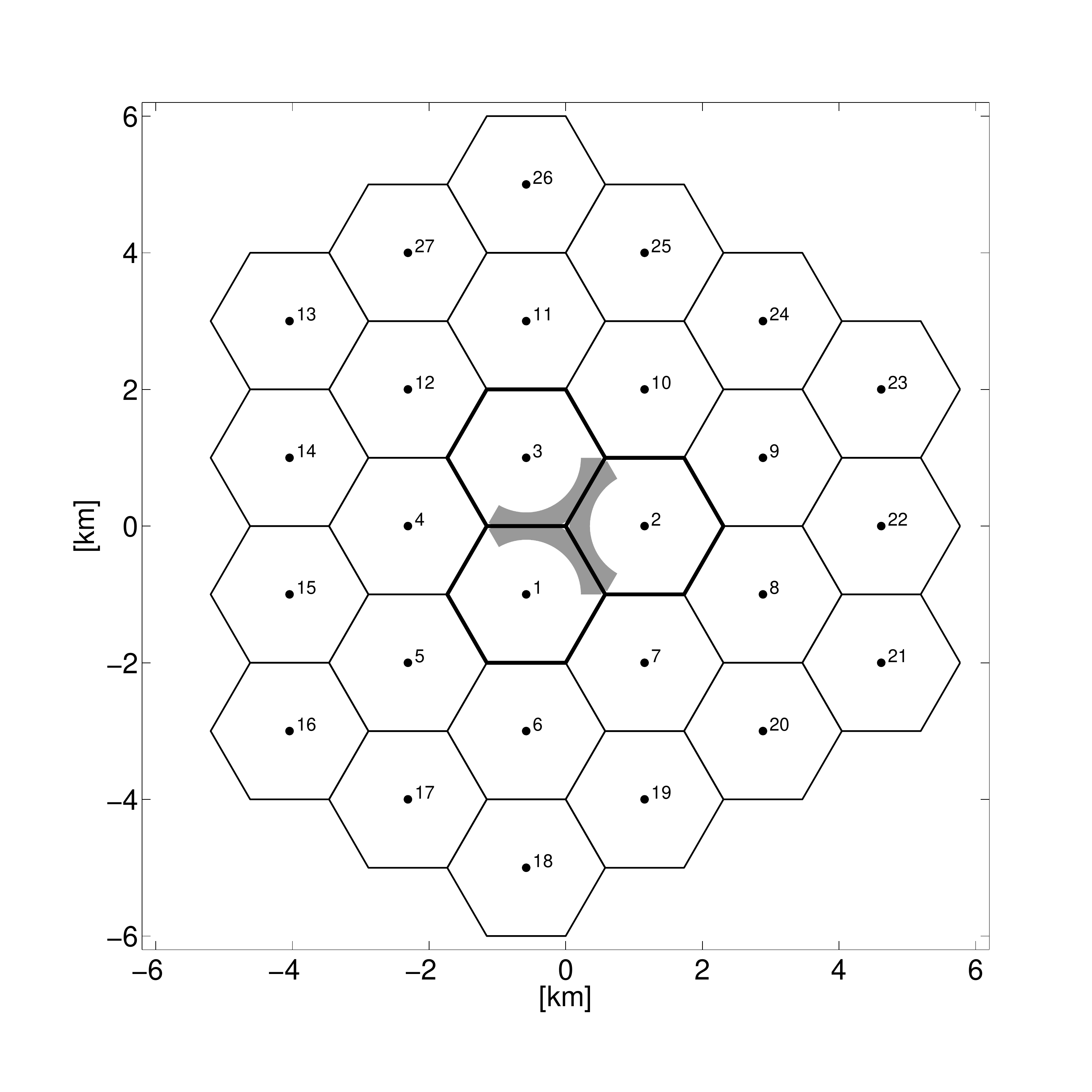}}
\vspace{-0.25cm}\caption{Simulated cellular network: BSs $1$, $2$, and $3$ are coordinated, and users are randomly dropped in the grey area.} \label{fig-network}
\end{figure}

In this section, we study the system performance via Monte-Carlo simulations. We consider the wireless cellular network in Figure\;\ref{fig-network}. BSs $1$, $2$, and $3$ coordinate their transmission (hence, $M=3$) on $N=16$ subcarriers with bandwidth $B=180$ kHz. Each BS serves three users, i.e., $|{\cal B}_m|=3$, which are uniformly distributed in the grey area. As to the power model, $\theta_1^{[n]}=0.25$ W,  $\theta_2^{[n]}=0.5$ W, $\theta_3^{[n]}=0.75$ W, and $\gamma_1^{[n]}=\gamma_2^{[n]}=\gamma_3^{[n]}=3.8$, for $n=1, \ldots, 16$, which are typical values for LTE systems  \cite{howmuch2011}. On each subcarrier, we consider Rayleigh fading, Log-Normal shadowing with standard deviation 8 dB, and the path-loss model  $\text{PL(d)}=\text{PL}_0\left(d_0/d\right)^{4}$, where $d\geq d_0$ is the distance in meters, and $\text{PL}_0$ is the free-space attenuation at the reference distance $d_0=100$ m with a carrier frequency of $1800$ MHz \cite{Goldsmith_book}.

Following \cite{fundamentalcooperation}, the noise variance ${\cal N}_s^{[n]}$ (which accounts for the power of both the thermal noise and the out-of-cluster interference) is modeled as
\begin{equation*}
{\cal N}_s^{[n]}=\underbrace{F{\cal N}_0 B}_{\text{thermal noise}}+ \underbrace{{P}_\text{out} \text{PL}_0
\sum_{j\in {\cal I}} \left( \frac{d_0}{d_{j,s}}\right)^4
\xi_{j,s}^{[n]}}_{\text{out-of-cluster interference}}
\end{equation*}
where $F=3$ dB is the noise figure of the receiver, ${\cal N}_0=-174$ dBm/Hz is the power spectral density of the thermal noise, ${\cal I}=\{4,5,\ldots, 27\}$ is the set of uncoordinated BSs in Figure\;\ref{fig-network}, ${P}_\text{out}$ is the average power radiated by the uncoordinated BSs on each subcarrier, $d_{j,s}$ is the distance from BS $j$ to user $s$, and $\xi_{j,s}^{[n]}$ is the Log-Normal shadowing (we assume that users only track long-term interference levels from uncoordinated BSs and, hence, short-term fading is averaged out). Notice that ${P}_\text{out}=0$ corresponds to the case in which the cluster of coordinated BSs is isolated.


The following analysis refers to a per-subcarrier power constraint with ${P}_{m,\max}^{[n]}={P}_{\max}/N$; a per-BS power constraint showed in our experiments a similar behavior and, hence, is not illustrated for brevity. Unless otherwise stated, the weights $w^{[n]}_{m, k(m,n)}$ in \eqref{eq:4} and \eqref{eq:prodEE} are set to $1/(MN)$. Finally, all plots are obtained after averaging over 1000 independent user drops.

\subsection{Implementation of the proposed algorithms}
All algorithms are initialized by assuming that the BSs transmit at the maximum power on each subcarrier. Moreover, letting $f_{\ell}$ be the value of the objective function at iteration $\ell$. The loop is stopped if $|f_{\ell}-f_{\ell-1}|/f_{\ell-1}<10^{-4}$ or a maximum number of 50 iterations has been reached.

The resource allocation strategies maximizing GEE, Sum-EE, and Prod-EE are referred to as GEE-opt, Sum-EE-opt, and Prod-EE-opt, respectively. For the sake of comparison, we also show the performance obtained by transmitting at the maximum power and by the resource allocation strategy in \cite{venturino-2009} maximizing the network sum-rate, i.e., the numerator of GEE in (\ref{eq:GEE}), which is referred to as sum-rate-opt.

\begin{figure}[!t]
\centerline{\includegraphics[width=0.95\columnwidth]{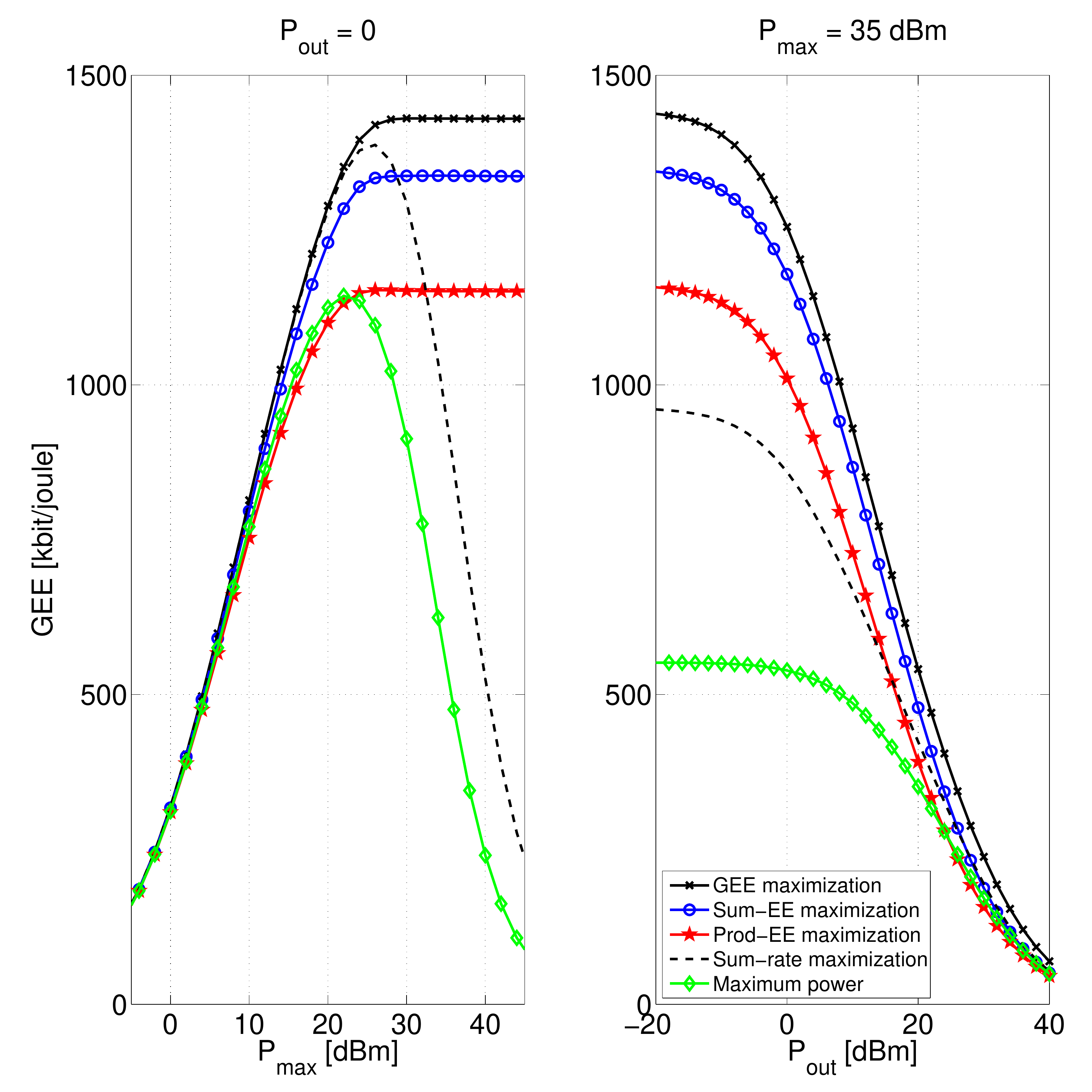}}
\vspace{-0.25cm}\caption{GEE vs $P_{\max}$ for ${P}_{\text{out}}=0$ (left); GEE vs ${P}_{\text{out}}$ for  $P_{\max}= 35$ dBm (right).} \label{fig-gee}
\end{figure}
\begin{figure}[!t]
\centerline{\includegraphics[width=0.95\columnwidth]{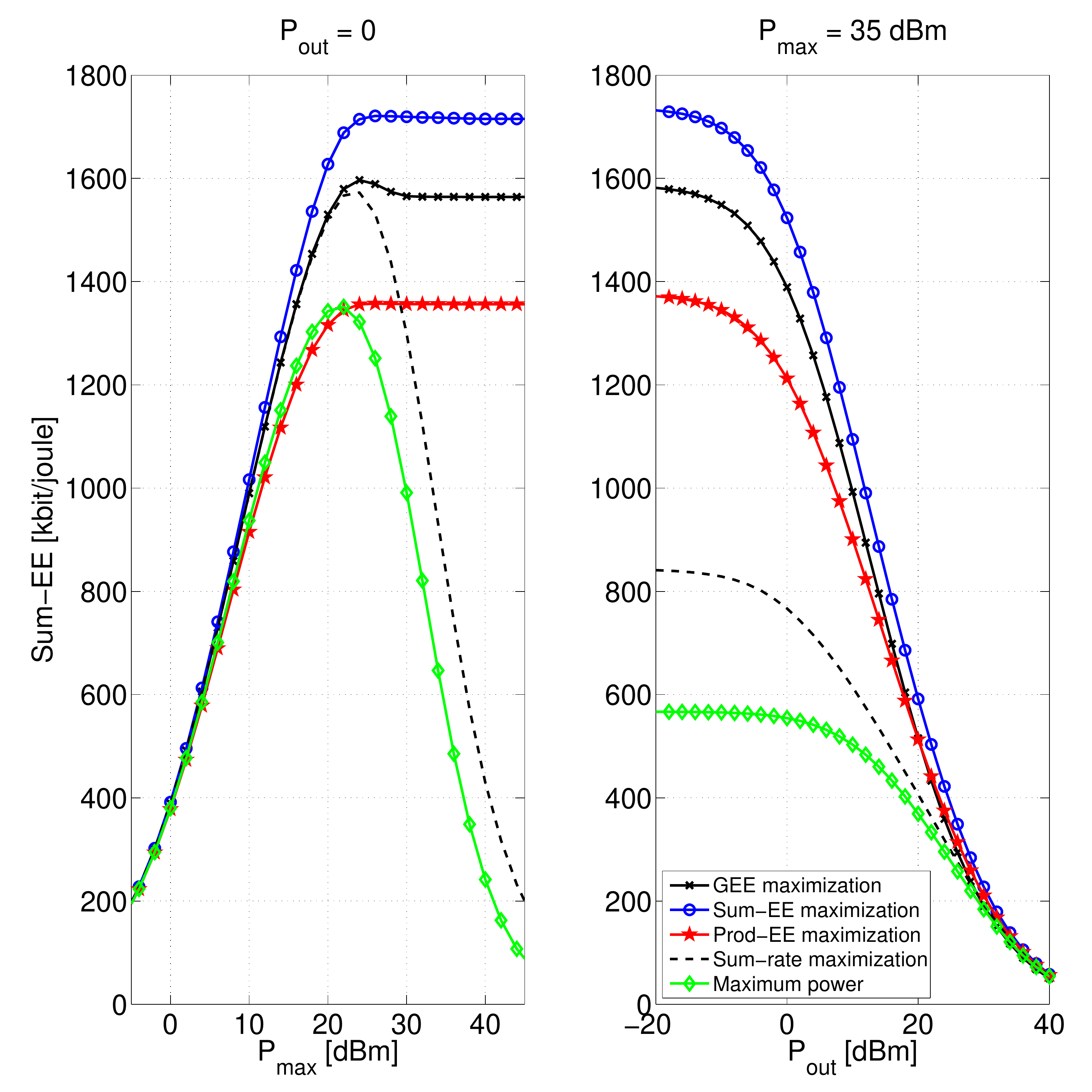}}
\vspace{-0.25cm}\caption{Sum-EE vs $P_{\max}$ for ${P}_{\text{out}}=0$ (left); Sum-EE vs ${P}_{\text{out}}$ for  $P_{\max}= 35$ dBm (right).} \label{fig-sumee}
\end{figure}
\begin{figure}[!t]
\centerline{\includegraphics[width=0.95\columnwidth]{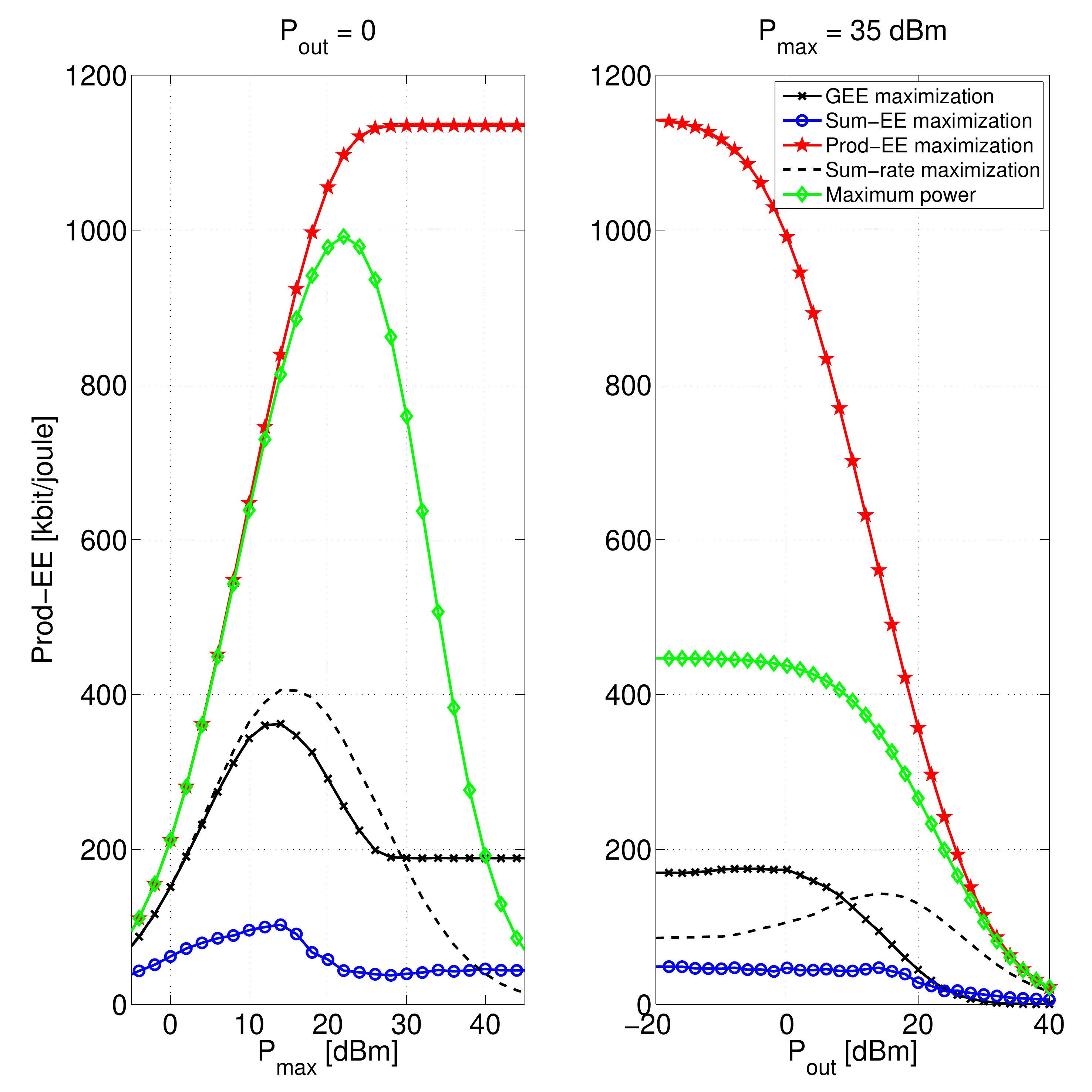}}
\vspace{-0.25cm}\caption{Prod-EE vs $P_{\max}$ for ${P}_{\text{out}}=0$ (left); Prod-EE vs ${P}_{\text{out}}$ for $P_{\max}= 35$ dBm (right).} \label{fig-prodee}
\end{figure}
\begin{figure}[!t]
\centerline{\includegraphics[width=0.95\columnwidth]{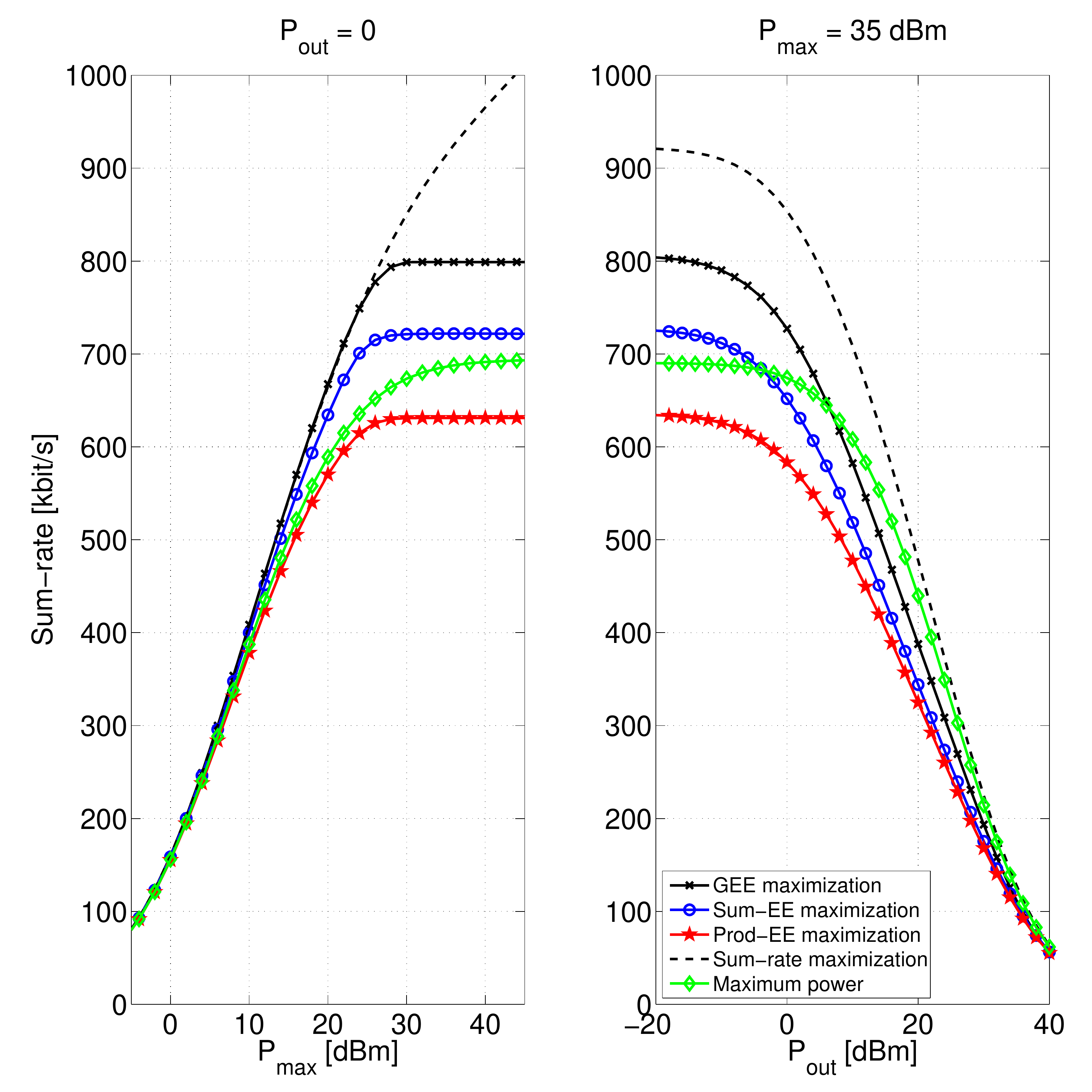}}
\vspace{-0.25cm}\caption{Sum-rate vs $P_{\max}$ for ${P}_{\text{out}}=0$ (left); sum-rate vs ${P}_{\text{out}}$ for $P_{\max}= 35$ dBm (right).} \label{fig-sumrate}
\end{figure}
\begin{figure}[!t]
\centerline{\includegraphics[width=0.95\columnwidth]{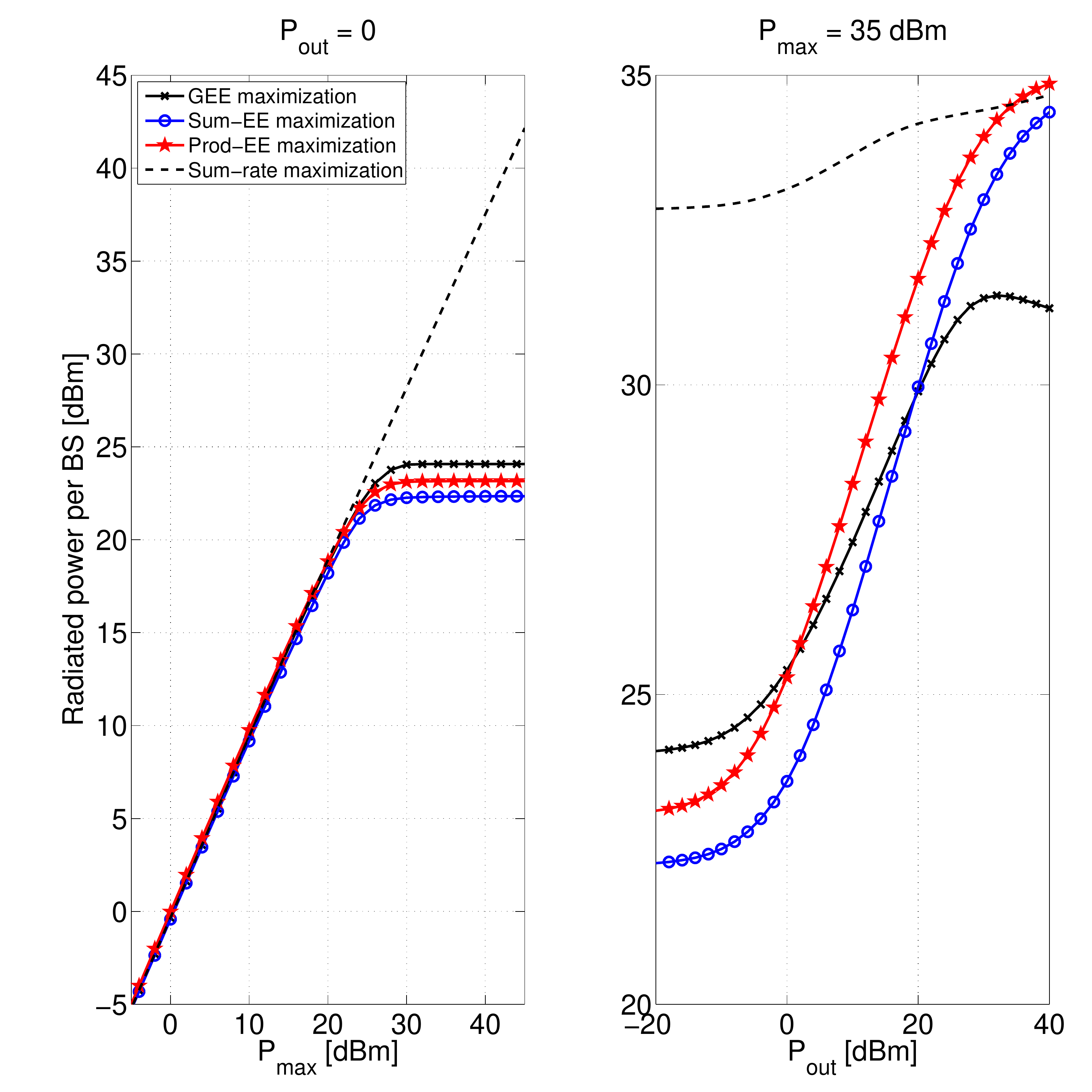}}
\vspace{-0.25cm}\caption{Per-BS radiated power vs $P_{\max}$ for ${P}_{\text{out}}=0$ (left); per-BS radiated power vs ${P}_{\text{out}}$ for $P_{\max}= 35$ dBm (right).} \label{fig-radiated}
\end{figure}
\begin{figure}[!tbp]
\centerline{\includegraphics[width=0.95\columnwidth]{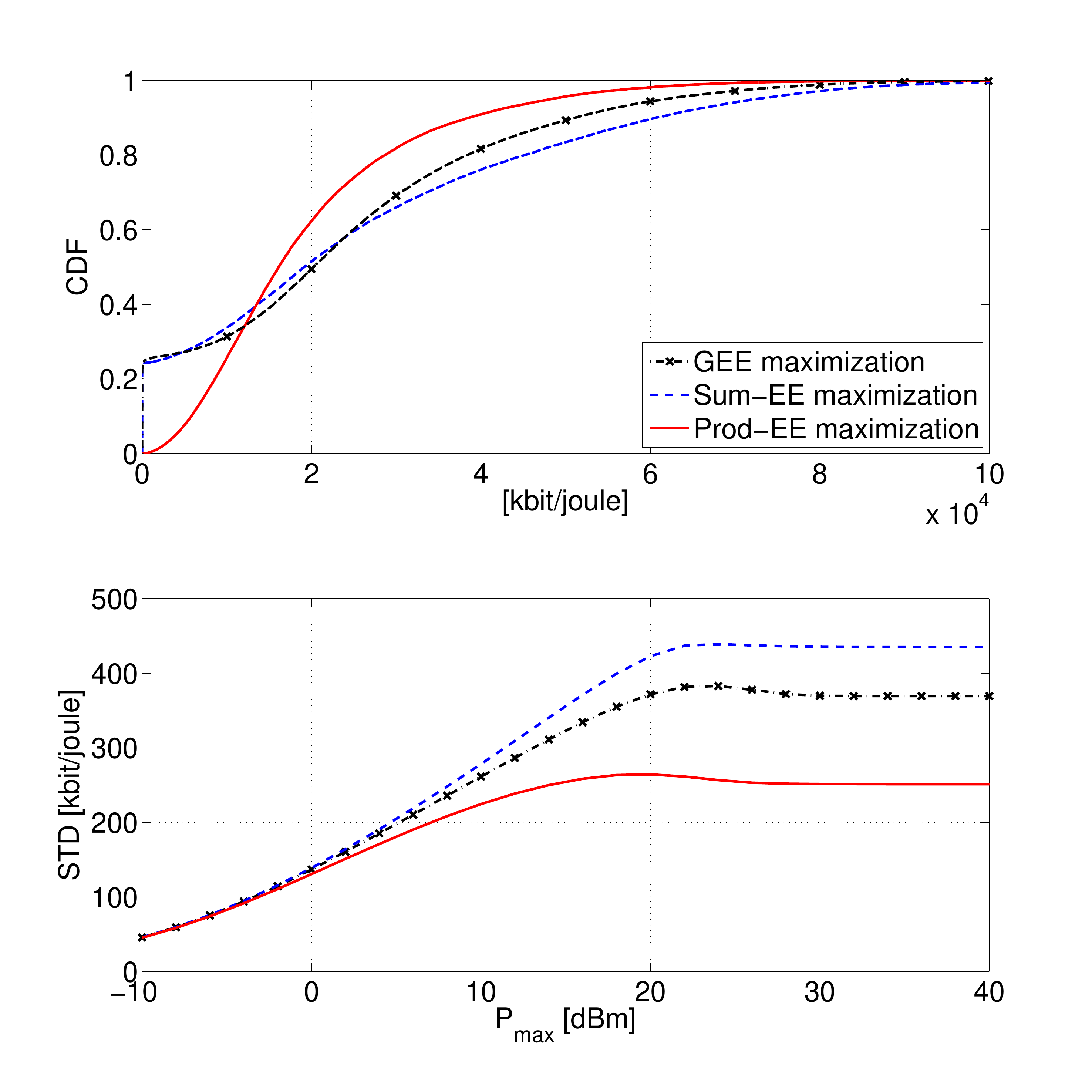}}
\vspace{-0.25cm}\caption{Empirical CDF of the energy efficiency achieved on each subcarrier when $P_{\max}=20$ dBm and $P_\text{out}=0$ (top); standard deviation of the energy efficiency achieved on each subcarrier versus $P_{\max}$ when $P_\text{out}=0$ (bottom).} \label{fig-fairness}
\end{figure}

\subsection{Performance results}

Figures~\ref{fig-gee}--\ref{fig-radiated} show GEE, Sum-EE, Prod-EE, sum-rate, and the  power radiated by each BS, respectively, for all considered resource allocations. In each figure, the subplot on the left refers to an isolated cluster, and the results are shown versus $P_{\max}$; the subplot on the right refers to a non-isolated cluster, and the results are shown versus ${P}_\text{out}$ for $P_{\max}=35$ dBm. For an isolated cluster, all solutions provide similar performance when $P_{\max}\leq 10$ dBm, since the radiated power consumption is negligible with respect to the static power consumption and, also, the cochannel interference is small compared to the noise power. For larger values of $P_{\max}$, instead, the considered figures of merit lead to different resource allocation strategies and, consequently, system performance. In this regime, the sum-rate-opt solution increases the network sum-rate  at the price of a heavy degradation in the system energy efficiency, no matter which definition of energy efficiency is considered (GEE, Sum-EE, or Prod-EE).  On the other hand, the GEE-opt, Sum-EE-opt, and Prod-EE-opt solutions exhibit a floor as $P_{\max}$ increases, since they do not use the excess available power to further increase the rate, as shown by Figure~\ref{fig-radiated}. For a non-isolated cluster, the value of GEE, Sum-EE, Prod-EE, and sum-rate degrade for increasing values of the out-of-cluster interference, irrespectively of the considered optimization criterion. Also, the performance gap among the considered solutions reduces as ${P}_\text{out}$ increases, since the out-of-cluster interference becomes dominant, making coordinated resource allocation less and less beneficial. It is interesting to notice that, in order to counteract the increased interference level, GEE-opt, Sum-EE-opt, and Prod-EE-opt solutions progressively use a larger fraction of the available power.



Figure~\ref{fig-fairness} shows the empirical CDF of the energy efficiency achieved on each subcarrier  for $P_{\max}=20$ dBm (top) and the standard deviation of the energy efficiency achieved on each subcarrier versus $P_{\max}$ (bottom), for the GEE-opt, Sum-EE-opt, and Prod-EE-opt solutions; an isolated cluster is considered. Results show that the energy efficiency achieved on the individual subcarriers is less dispersed for the Prod-EE-opt allocation, thus confirming that Prod-EE maximization provides a more balanced use of the available subcarriers.

\begin{figure}[!tbp]
\centerline{\includegraphics[width=0.95\columnwidth]{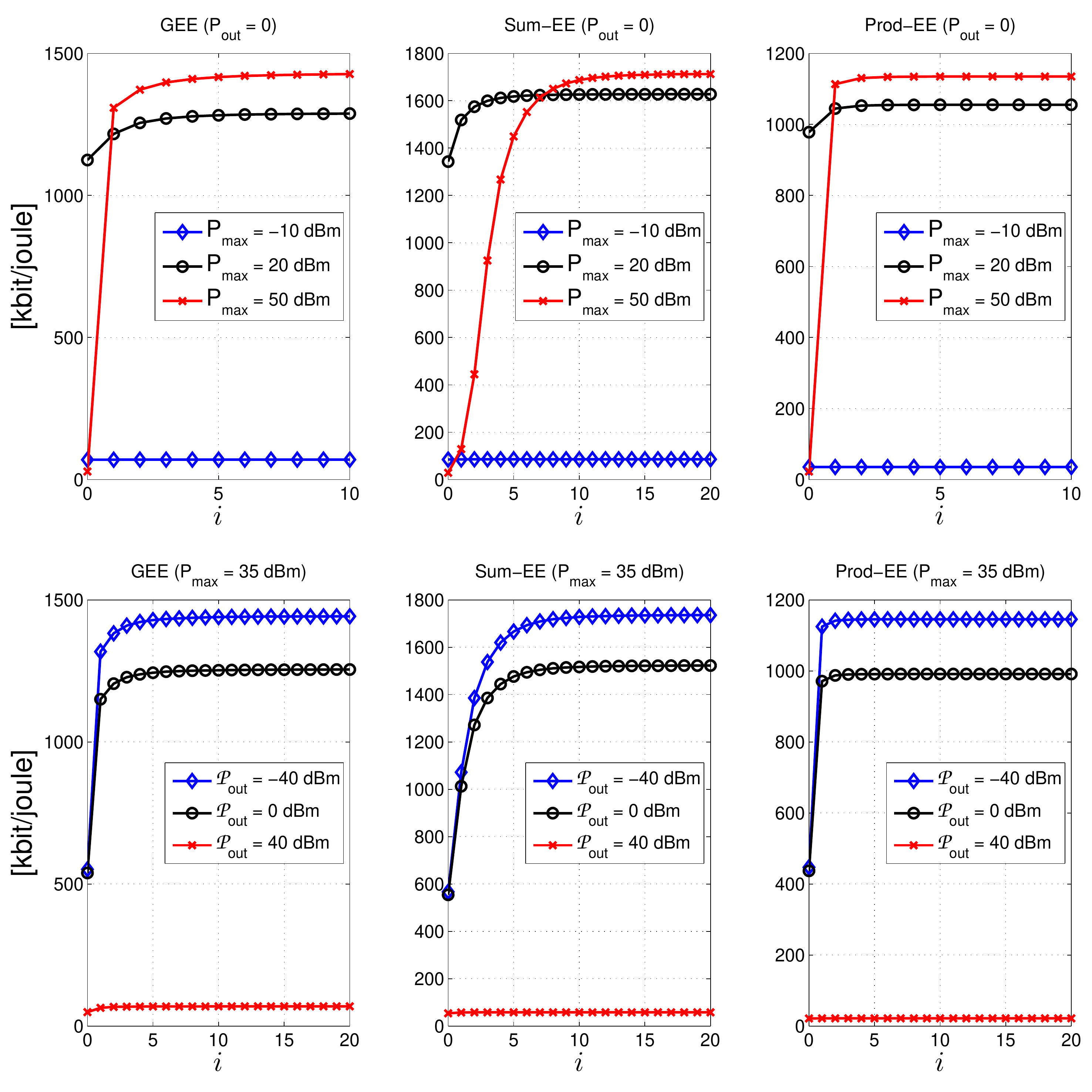}}
\vspace{-0.25cm}\caption{GEE, Sum-EE, and Prod-EE versus the number $i$ of iterations of Algorithms~\ref{alg-Papandriopoulus}, \ref{alg-2}, and~\ref{alg-ProdLog1}, respectively. Top: $P_{\max}=-10,20,50$ dBm and ${P}_{\text{out}}=0$. Bottom: ${P}_{\text{out}}=-40,0,40$ dBm and $P_{\max}=35$ dBm.} \label{fig-convergence1}
\end{figure}

Finally, we study the convergence of the proposed algorithms. Figure~\ref{fig-convergence1} reports the value of GEE, Sum-EE, and Prod-EE versus the number $i$ of iterations of Algorithms~\ref{alg-Papandriopoulus}, \ref{alg-2}, and~\ref{alg-ProdLog1}, respectively. The upper plots refer to an isolated cluster, while the lower plots to a non-isolated cluster. All algorithms reach a steady value in few iterations in all considered scenarios; also, the convergence speed decreases for increasing values of
$P_{\max}$ and for diminishing values of ${P}_\text{out}$, i.e., when the system performance is limited by the co-channel interference generated by the other coordinated BSs.

\subsection{Influence of the weights}
\begin{figure}[!tbp]
\centerline{\includegraphics[width=0.95\columnwidth]{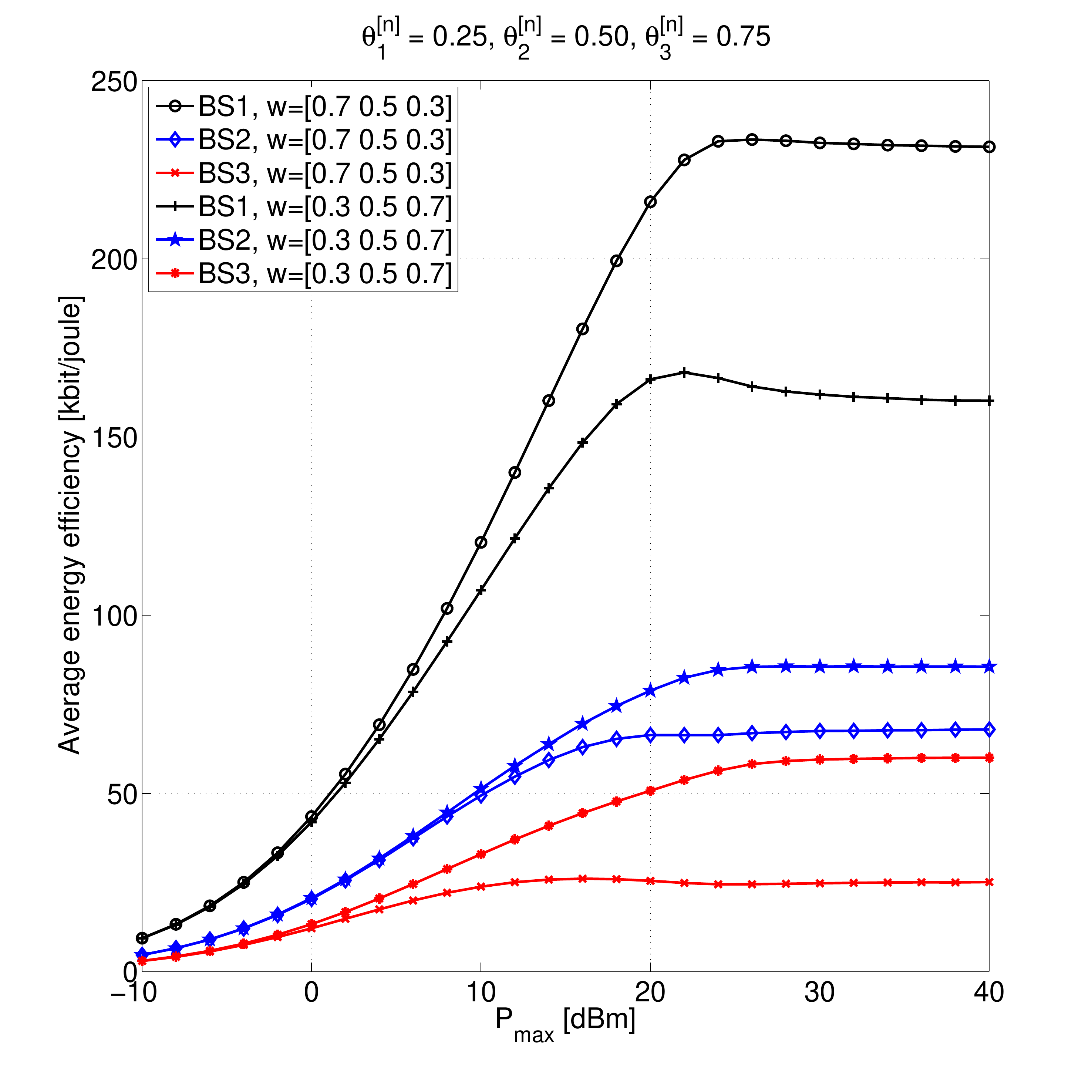}}
\vspace{-0.25cm}\caption{Average energy efficiency of each coordinated BS versus $P_{\max}$. The Sum-EE-opt solution and an isolated cluster are considered.} \label{fig-weights}
\end{figure}

The weights in the definition of Sum-EE and Prod-EE may be used to give priority to specific subcarriers and/or BSs; this is an attractive feature, especially in heterogeneous scenarios. As an  example, we consider the maximization of Sum-EE with two choices of the weights: a) $w_{1,s}^{[n]}=0.7$, $w_{2,s}^{[n]}=0.5$, and $w_{3,s}^{[n]}=0.3$;  b) $w_{1,s}^{[n]}=0.3$, $w_{2,s}^{[n]}=0.5$, and $w_{3,s}^{[n]}=0.7$. In Figure~\ref{fig-weights}, we consider the Sum-EE-opt solution and report the average energy efficiency of each coordinated BS, i.e.,
\begin{equation*} \label{eq:avee-perBS}
\frac{1}{N}\sum_{n=1}^{N}\frac{R_{m,k(m,n)}^{[n]}}{\gamma_{m}^{[n]}p_{m}^{[n]}+\theta_{m}^{[n]}}
\end{equation*}
for $m=1,\ldots,M$. An isolated cluster is considered, and the results are plotted versus $P_{\max}$.
Since $\theta_{1}^{[n]}=0.25$, $\theta_{2}^{[n]}=0.5$, and $\theta_{3}^{[n]}=0.75$, BS1 is the most energy-efficient BS, while BS3 is the most energy-inefficient BS.
In the first scenario, BS1 achieves an average energy-efficiency much larger than that of other BSs, as it has the largest priority and the best energy efficiency. Instead, BS3 is extremely penalized, as it has the worst energy efficiency and the smallest priority. In the second scenario, a more balanced resource allocation is obtained by assigning a higher priority to BS3 and a lower priority to BS1. Notice that the performance of BS2 remains approximatively unchanged, as its weights are kept fixed.

\subsection{Impact of the number of subcarriers and users.}

Experiments have been carried out to also study the impact of the number of subcarriers and users on the system performance; we offer here some general comments on the results, without including any detailed plot for the sake of brevity.

If the maximum transmit power scales linearly with $N$, the system performance are only marginally affected by the number of subcarriers.
To be more specific, let us first focus on GEE. When $N$ scales up, the system sum-rate proportionally increases, as more subcarriers are available for transmission and the average transmit power per-subcarrier remains fixed. At the same time, the power consumption is also increased, and the ratio between the sum-rate and sum-power remains substantially unchanged.
Similarly, the energy efficiency of each individual subcarrier remains approximatively constant, and, consequently,  the arithmetic and geometric means of the energy efficiencies across all subcarriers do not scale with $N$.

In keeping with intuition, if the number of users in the coordinated cluster is increased, the system performance tends to improve due to the multiuser diversity gain \cite{Goldsmith_book}. Clearly, while the network-wide performance improves, the average physical resources assigned to each user progressively reduce.

\subsection{Discussion on algorithms' complexity}
As seen from Figure~\ref{fig-convergence1}, the proposed algorithms converge in only 5 - 15 iterations, depending on the operating scenario. For each method, the complexity of a single iteration is mainly tied to the optimization of the transmit power.

In Algorithm~\ref{alg-Papandriopoulus}, the power update requires the solution of the fractional program (\ref{eq:problem-LB-Papandriopoulus}). Several equivalent methods exist to tackle fractional problems \cite{Isheden2011}, and Algorithm \ref{alg-Papandriopoulus} is independent of the employed method.
Here, we have resorted to Dinkelbach's algorithm, which is widely-used in the literature. The Dinkelbach's algorithm has a super-linear convergence rate \cite{Isheden2011} and, in each iteration, only requires the solution of a convex problem, which can be accomplished in polynomial time by means of many convex programming algorithms \cite{Boyd_book}. In Algorithm~\ref{alg-2}, the power update just requires the computation of the algebraic expressions in (\ref{eq:12}) and the solution to the waterfilling-like problems in  \eqref{eq:modified-WF-2}, which can be accomplished in logarithmic time through a bisection search.
Finally,  the power update  in Algorithm~\ref{alg-ProdLog1}  requires the solution of a convex problem, which again is accomplished in polynomial time. 

Complexity generally grows with the number of coordinated BSs and active users. However, the number of coordinated BSs is usually in the order of few units, since the advantages of cooperation with far-away BSs are marginal. Moreover, coordination may only be performed for cell-edge users, which typically do not experience favorable propagation conditions. 

\section{Conclusions}\label{SEC:Conclusions}
We have studied the problem of resource allocation in the downlink of an OFDMA network with base station coordination.  Three figures of merit have been considered for system design, namely, the ratio of the network sum-rate to the network power consumption (GEE), the weighted sum of the energy efficiencies on each subcarrier (Sum-EE), and the exponentially-weighted product of the energy efficiencies on each subcarrier (Prod-EE). Algorithms for coordinated user scheduling and power allocation have been proposed, under a per-BS or per-subcarrier power constraint. In particular, GEE is optimized by solving a series of concave-convex fractional relaxations, while for Prod-EE a series of concave relaxations is considered; as to Sum-EE, an iterative method to solve the Karush–Kuhn–Tucker conditions is proposed. For all figures of merit, algorithms to compute a globally-optimal solution are derived in the asymptotic noise-limited regime. 
It has been shown that Sum-EE and Prod-EE provide more degrees of freedom for system design compared to the more popular GEE, as the corresponding weights may be used to give priority to specific subcarriers and/or base stations. Also, Prod-EE inherently makes a more balanced use of the available spectrum, preventing the unpleasant situation where few subcarriers receive most of the system resources.

\appendix
\subsection{Proof of Lemma~\ref{Lemma-1}}\label{Proof-Lemma-1}
Since the bound in (\ref{eq:LB-Papandriopoulus}) is tight at $\text{SINR}_{m, k(m,n)}^{[n]}=\bar{z}_m^{[n]}$, with $\bar{z}_m^{[n]}\geq 0$, we have
\begin{equation*}
\max_{\mathbf{q}\in {\cal Q}} f(\exp\{\mathbf{q}\},\mathbf{k})\geq B\sum_{m=1}^{M}\sum_{n=1}^{N} \log_2\left(1+\bar{z}_m^{[n]}\right) \geq 0.
\end{equation*}
Finally, concavity of $f(\exp\{\mathbf{q}\},\mathbf{k})$ and convexity of $g(\exp\{\mathbf{q}\})$ follow from the fact that the log-sum-exp function is convex \cite{Boyd_book}.

\subsection{Proof of Proposition~\ref{Prop-1}}\label{Proof-Prop-1}
Let $\mathbf{q}_{\ell}=\ln\mathbf{p}_{\ell}$ and $\mathbf{k}_{\ell}$ be the optimized values
after $\ell$ iterations. Also, let $h_{\ell}$ be the lower bound in (\ref{eq:LB-Papandriopoulus}) when the approximation constants are computed according to $\mathbf{p}_{\ell}$ and $\mathbf{k}_{\ell}$.  Then,  we have:
\begin{eqnarray*}
\ldots&\overset{(d)}{\leq}& \text{GEE}(\mathbf{p}_{\ell},\mathbf{k}_{\ell})\overset{(a)}{=}h_{\ell}(\mathbf{q}_{\ell},\mathbf{k}_{\ell})\\
&\overset{(b)}{\leq}& h_{\ell}(\mathbf{q}_{\ell+1},\mathbf{k}_{\ell})\overset{(c)}{\leq}  \text{GEE}(\mathbf{p}_{\ell+1},\mathbf{k}_{\ell})\\
&\overset{(d)}{\leq}& \text{GEE}(\mathbf{p}_{\ell+1},\mathbf{k}_{\ell+1})\overset{(a)}{=}h_{\ell+1}(\mathbf{q}_{\ell+1},\mathbf{k}_{\ell+1})\overset{(b)}{\leq} \ldots
\end{eqnarray*}
where the equality (a) is due to the fact that the relaxation (\ref{eq:LB-Papandriopoulus}) is tight at the current SINR values; the inequality (b) is due to the fact that the Dinkelbach's procedure computes the globally-optimal solution to \eqref{eq:problem-LB-Papandriopoulus}; the inequality (c) follows from (\ref{eq:LB-Papandriopoulus}); the inequality (d) follows from the fact that the user selection in (\ref{eq:13-LB}) does not decrease the value of
GEE. Since GEE is bounded above, the procedure must converge.

Next, the KKT conditions for (\ref{eq:problem-LB-Papandriopoulus}) can be written as
\begin{equation}\label{eq:KKT-LB-Papandriopoulus}
\begin{array}{l}\displaystyle
\frac{\text{d}}{\displaystyle \text{d} q_m^{[n]}}h(\exp\{\mathbf{q}\},\mathbf{k})
-\lambda_{m}\exp\{q_{m}^{[n]}\}=0,\;\forall\;m,n\\
\lambda_{m}\geq0, \;\forall\;m\\
\sum_{n=1}^{N}\exp\{q_m^{[n]}\}\leq P_{m,\max},\;\forall\;m\\
\lambda_{m}\left(P_{m,\max}-\sum_{n=1}^{N}\exp\{q_m^{[n]}\}\right)=0,\;\forall\;m
\end{array}
\end{equation}
where $\lambda_{m}$ is the Lagrange multiplier with respect to the power constraint of BS $m$ and
\begin{multline}\notag
\displaystyle \frac{\text{d}}{\displaystyle \text{d} q_m^{[n]}}h(\exp\{\mathbf{q}\},\mathbf{k}) =
\dfrac{B/\ln2}{g(\exp\{\mathbf{q}\})}\times  \\
\left[\alpha_{m}^{[n]}-\displaystyle \sum_{j=1,\;j\neq m}^{M}
\dfrac{\alpha_{j}^{[n]}\exp\{q_m^{[n]}\}G_{m,k(j,n)}^{[n]}}{\displaystyle 1+\sum_{\ell=1,\;\ell\neq j}^{M}\exp\{q_{\ell}^{[n]}\}G_{\ell,k(j,n)}^{[n]}}\right] -\notag \\
\gamma_{m}^{[n]}\exp\{q_m^{[n]}\}\dfrac{h(\exp\{\mathbf{q}\},\mathbf{k})}{g(\exp\{\mathbf{q}\})}.
\end{multline}
Let $(\bar{\mathbf{q}},\bar{\mathbf{k}})$ be the solution provided by Algorithm~\ref{alg-Papandriopoulus} upon convergence. Then, there exists a set of multipliers
$\bar{\bm{\lambda}}$ such that the triplet $(\bar{\mathbf{q}},\bar{\mathbf{k}},\bar{\bm{\lambda}})$ simultaneously satisfies \eqref{eq:13-LB} and \eqref{eq:KKT-LB-Papandriopoulus}.

Finally, notice that the first-order optimality conditions for (\ref{eq:5A_LB}) in the $\mathbf{q}$-space are  given by (\ref{eq:13-LB}) and
\begin{equation}\label{eq:KKT-GEE}
\begin{array}{l}\displaystyle
\frac{\text{d}}{\displaystyle \text{d} q_m^{[n]}}\text{GEE}(\exp\{\mathbf{q}\},\mathbf{k})
-\lambda_{m}\exp\{q_m^{[n]}\}=0,\;\forall\;m,n\\
\lambda_{m}\geq0, \;\forall\;m\\
\sum_{n=1}^{N}\exp\{q_m^{[n]}\}\leq P_{m,\max},\;\forall\;m\\
\lambda_{m}\left(P_{m,\max}-\sum_{n=1}^{N}\exp\{q_m^{[n]}\}\right)=0,\;\forall\;m
\end{array}
\end{equation}
where
\begin{multline}\notag
\displaystyle
\frac{\text{d}}{\displaystyle \text{d} q_m^{[n]}}\text{GEE}(\exp\{\mathbf{q}\},\mathbf{k}) =\dfrac{B/\ln2}{g(\exp\{\mathbf{q}\})}\times \notag \\
\left[\displaystyle\frac{\exp\{q_m^{[n]}\} G_{m,k(m,n)}^{[n]} }{\displaystyle 1+\sum_{\ell=1}^{M}\exp\{q_{\ell}^{[n]}\} G_{\ell,k(m,n)}^{[n]}}
\right. \\
 \left.-\displaystyle\sum_{j=1,\;j\neq m}^{M}
\dfrac{\text{SINR}_{j,k(j,n)}^{[n]}\exp\{q_m^{[n]}\}G_{m,k(j,n)}^{[n]}
}{\displaystyle 1+\sum_{\ell=1}^{M}\exp\{q_{\ell}^{[n]}\}G_{\ell,k(j,n)}^{[n]}}\right]-\notag \\
\gamma_{m}^{[n]}\exp\{q_m^{[n]}\}
\dfrac{\text{GEE}(\exp\{\mathbf{q}\},\mathbf{k})}{g(\exp\{\mathbf{q}\})}.
\end{multline}
The proof is completed by noticing that the approximation in (\ref{eq:LB-Papandriopoulus}) is exact at convergence and, therefore, the triplet $(\bar{\mathbf{q}},\bar{\mathbf{k}},\bar{\bm{\lambda}})$ also solves \eqref{eq:13-LB} and \eqref{eq:KKT-GEE}.

\subsection{Proof of Proposition~\ref{Prop-1-NL}}\label{Proof-Prop-1-NL}
Notice that $\text{GEE-NL}(\mathbf{p},\mathbf{k})$ is a strictly pseudo-concave function of $\mathbf{p}$, since it is the ratio between a strictly concave function and a linear function \cite{Avriel-book}. This implies that the optimal resource allocation strategy  can be found by alternatively computing the best $\mathbf{k}$ and $\mathbf{p}$, as summarized in Algorithm~\ref{alg-Papandriopoulus-NL}.

\subsection{Proof of Lemma~\ref{Prop:RegionConcave}}\label{Proof-Lemma-2}
Since $u$ is strictly pseudo-concave, $\bar{x}$ is the unique solution of the equation
\begin{equation}\label{Eq:Der1}
\frac{du(x)}{dx}=0 \leftrightarrow \frac{a(x+c)}{1+ax}=\frac{\log_{2}(1+ax)}{\log_{2}e}\;.
\end{equation}
Moreover, for $x\leq \bar{x}$, the left hand side (LHS) of (\ref{Eq:Der1}) is larger or equal than the RHS, whereas for $x> \bar{x}$, the RHS is larger than the LHS.
Next, all points in the concave region of $u$ must satisfy the following condition \cite{Boyd_book}
\begin{equation*}\label{Eq:Der2}
\frac{d^{2}u(x)}{dx^{2}}\leq 0 \leftrightarrow \frac{a(x+c)}{1+ax}+\frac{a^{2}(x+c)^{2}}{2(1+ax)^{2}}\geq\frac{\log_{2}(1+ax)}{\log_{2}e}.
\end{equation*}
Since $\frac{a^{2}(x+c)^{2}}{2(1+ax)^{2}}>0$, the last inequality holds at least for all $0\leq x\leq \bar{x}$.

\bibliographystyle{IEEEtran}

\bibliography{IEEEfull,references}
\end{document}